\documentclass[journal]{IEEEtran}
\ifCLASSINFOpdf
\else
\fi
\usepackage{multicol}

\usepackage{amssymb}
\usepackage{amsmath}

\newtheorem{theorem}{\textbf{Theorem}}

\usepackage{algorithm}
\usepackage{algorithmicx}
\usepackage{algpseudocode}

\usepackage{array}

\usepackage[backref]{hyperref}
\usepackage{url}

\usepackage{graphicx}

\makeatletter
\renewcommand{\maketag@@@}[1]{\hbox{\m@th\normalsize\normalfont#1}}%
\makeatother

\begin{document}
%
\title{Generalized Multi-cluster Game under Partial-decision Information  with Applications \\ to Management  of Energy Internet  }
%
%
%

\author{
Yue Chen and  Peng Yi	
\thanks{The paper was sponsored by
	the National Natural Science Foundation of China under Grant No. 62003239. \textit{(Corresponding author: Peng Yi.)}}

\thanks{The authors are with the Department of Control Science and Engineering,
	Tongji University, Shanghai, 201804, China, and also with  Shanghai Institute of Intelligent Science and Technology, Tongji University, Shanghai 200092, China. (email: chenyue\_j@tongji.edu.cn, yipeng@tongji.edu.cn).  }
}

\maketitle

\begin{abstract}
The decision making and management of many engineering networks involves multiple parties with conflicting interests, while each party is constituted with multiple agents. Such problems can be casted as a multi-cluster game.
Each cluster is treated as a self-interested  player in a non-cooperative game where agents in the same cluster cooperate together to
optimize the payoff function of the cluster.
In a large-scale network, the information of agents in a cluster can not be available immediately for agents beyond this cluster, which raise challenges to the existing Nash equilibrium seeking algorithms.
Hence, we consider a partial-decision information scenario in generalized Nash equilibrium  seeking for multi-cluster games in a distributed manner.
We reformulate the problem as finding zeros of the sum of preconditioned monotone operators by the primal-dual analysis and graph Laplacian matrix.
Then a distributed generalized Nash equilibrium seeking algorithm is proposed without requiring fully awareness of its opponent clusters' decisions based on a forward-backward-forward method.
With the algorithm, each agent estimates the strategies of all the other clusters by communicating with neighbors via an undirected network.
We show that the derived operators can be monotone when the communication strength parameter is sufficiently large. We prove the algorithm convergence resorting to the fixed point theory by providing a sufficient condition.
We discuss its potential application in Energy Internet with numerical studies.
\end{abstract}


%
\IEEEpeerreviewmaketitle

\section{Introduction}
%
%
%
%
\IEEEPARstart{N}ASH game plays an important role in engineering network systems with multiple agents, where each agent aims to minimize its own payoff function interfered by other player's decisions.
It arises in a broad variety of applications, such as
transportation network \cite{hollander2006applicability},
wireless network \cite{charilas2010survey},
construction management \cite{kaplinski2010game},
mobile sensor network \cite{stankovic2011distributed}.
In this paper, we consider a game composed of multiple clusters (or coalitions \cite{ye2018nash}, \cite{zeng2017distributed}) while each cluster is treated as a non-cooperative player in the game,
and the agents within the same cluster work collaboratively to minimize the cluster's payoff function.
This formulation is discussed in \cite{ye2018nash}, \cite{ye2019unified}, and Nash equilibrium (NE) seeking algorithms are designed.
It is an extension of the zero-sum games of two networks in \cite{gharesifard2013distributed}  and \cite{lou2015nash} by allowing more
clusters.
The multi-cluster game and its NE seeking have received considerable attention,
due to its wide applications in large-scale hierarchical networks, including healthcare networks \cite{peng2009coexistence},
task allocation networks \cite{shehory1998methods},
electricity market \cite{zeng2019}.
\par In some engineering problems, both the payoff function and the feasible action set of each player can be coupled with the other's actions, and the game is called generalized Nash game \cite{pang2008distributed}, \cite{zhu2016distributed}, \cite{ghaderi2014opinion}, \cite{deng2018distributed}.
Applications of	 such games are ubiquitous in the engineering field,
including smart grids \cite{deng2021distributed},
 optical network \cite{pavel2007extension},
mobile cloud computing \cite{cardellini2016game},
charging control of plug-in electric vehicle \cite{grammatico2017dynamic}
and management of energy hubs \cite{liang2019generalized}.
Generalized Nash equilibrium (GNE) is an solution to generalized Nash game, and its seeking algorithms have been studied in
\cite{pavel2007extension},
\cite{facchinei2007finite},
\cite{facchi2010GNE},
\cite{arslan2014games}.
GNE seeking for multi-cluster game are also studied, such as \cite{zeng2017distributed}, \cite{zeng2019}.
\par The classical NE seeking studies usually assume that all agents can observe the full information of all their opponents' decisions.
In this setting,  a central coordinator or an interference network is used to communicate with all the agents
\cite{swenson2015empirical}, \cite{zhu2016distributed}, while the NE computing method is usually based on best-response
\cite{facchinei2007finite},
 or gradient schemes
 \cite{lei2020synchronous}.
However, in recent years, there has been an increasing interest to consider partial-decision information regimes that agents are unable to fully observe opponents' decisions.
 In partial information regime, many studies consider consensus-based gradient type approach of NE seeking,
 e.g. \cite{zimmermann2021solving}, \cite{meng2020linear} and \cite{nian2021distributed} all use the gradient tracking play to calculate NE in multi-cluster games, 
 and \cite{pang2020gradient} designs a distributed NE seeking algorithm by using an inexact-ADMM approach.
 All the aforementioned literatures consider games (including multi-cluster games) without shared constraints, i.e., non-generalized games.
As far as we known, there are little literatures discussing generalized games among multiple clusters except \cite{zeng2019},
even fewer in partial information regime of multi-cluster games.
\par In this paper we consider distributed GNE computation in multi-cluster games with shared affine coupling constraints under partial information regime.
Solving the generalized Nash equilibrium problems is a challenging task,  and one useful way is to find an important class of GNEs called variational GNE (v-GNE) \cite{facchinei2007finite}, \cite{facchi2010GNE}, \cite{kulkarni2012variational}, which corresponds to the solution of a variational inequality (VI).

\par In recent years, operator-theoretic approach  gets appealing for solving (generalized) Nash games, since it is  regarded  being mathematically general and elegant.
The GNE solving problems are reformulated into finding zeros of monotone operators, known as monotone inclusion problems \cite{bauschke2011convex}.
 \cite{yi2019operator} proposed a popular forward-backward (FB) operator splitting method by using preconditioned matrix  and incorporating the Laplacian matrix of a connected graph to enforce the consensus of local multipliers,
 and  the convergence is guaranteed by the cocoercivity and monotonicity of the operators with fixed step-sizes.
Later on, \cite{pavel2019distributed} discusses the GNE seeking under partial decision information regime by introducing an auxiliary variable to guarantee the monotonicity of operator in FB method.
\cite{yi2019asynchronous} considers the same situation with delayed information.
In \cite{bianchi2022fast}, a proximal point method is utilized to deal with GNE under partial information setting.
Another fixed-point iteration approach called forward-backward-forward (FBF) is also employed in NE seeking since it relaxes the strong/strict monotonicity assumption for FB methods to monotonicity on the (extended) pseudo-gradient mapping of the game \cite{cui2021relaxed}, \cite{franci2020half}.
\par Motivated by the above, we aim to propose a distributed algorithm via a preconditioning FBF approach to solve the GNE in multi-cluster games
under partial-decision information regime.
With Karush-Kuhn-Tucker (KKT) conditions of the VI and a primal-dual analysis, we endow local copies of the multipliers and auxiliary variables for agents, noticing that the KKT condition requires the leader agent of each cluster to reach consensus on the multiplier of the shared constraint.
We reformulate the problem as finding zeros of a mapping corresponding to the fixed points of a certain operator by constructing an symmetric, positive definite matrix, known as preconditioning matrix.
The FBF method is employed to develop the distributed discrete-time algorithm, and the algorithm convergence is guaranteed with a  fixed step-size.
 The contributions of this paper can be summarized as follows:
\begin{enumerate}	
	\item A generalized game composed of multiple clusters is considered.
	Each cluster is a non-cooperative player where agents in the same cluster cooperate together to optimize the cluster's payoff function.
	The communication graph is hierarchical, and is constituted with leader-cluster graph and inner-cluster graph (called inter-cluster graph and intra-cluster graph in \cite{zeng2019}).
	\item
	The work considers the partial-decision information setting of GNE seeking for multi-cluster games.
    The proposed algorithm relaxes the (extended) pseudo-gradient mapping condition from  strong monotonicity to monotonicity.
	The algorithm is in discrete-time and can be easily implemented in practical engineering applications.
	And with the fixed step-size, the algorithm has faster convergence than algorithms with vanishing step-sizes.
	To deal with the loss of monotonicity caused by partial information, a communication strength parameter is designed to ensure the	restrict monotonicity of the operator.
	\item We discuss the application of multi-cluster game to Energy Internet(EI) \cite{chao2021hierarchical}, \cite{saad2012game}, which was aroused to be a feasible solution to energy crisis \cite{huang2010future}, \cite{sun2019energy}, \cite{wang2017survey}.
	Each energy subnet can provide storage energy to compete for the utility while satisfying the demands.
	The decision-making of storage energy export of each energy-subnet is formulated as a GNE seeking of multi-cluster game with the proposed algorithm.
\end{enumerate}

\par This paper is organized as follows.
Section \ref{section2} formulates the multi-cluster game with partial-decision information.
Section \ref{section3} introduces the distributed method  to the v-GNE seeking of  multi-cluster game.
Section \ref{section4} contains two subsections. The subsection \ref{subsubsectionA} introduces the development of the distributed algorithm based on operator splitting method, and subsection \ref{subsubsectionB} shows the convergence analysis.
Applications to EI is discussed in Section \ref{section5}, and concluding remarks are given in Section \ref{section6}.
The proofs are given in Appendix \ref{appendix}.
\par \textbf{Notations}. Denote $\mathbb{R}^m(\mathbb{R}^m_+)      $  as the $ m $-dimensional (non-negative) Euclidean space. For an integer $ m>0 $, denote $ \mathbb{N}_{m}^{*}:=\{1,2,...,m\} $. And $ \mathbb{N}_{m}^{*} \backslash j := \{1,2,\dots,j-1,j+1,\dots,m-1,m\} $, which contains all elements of set $  \mathbb{N}_{m}^{*} $ except $ j $.
$ \left   \langle x,y  \right \rangle= x^Ty $ denotes the inner product of $ x $, $ y $, and $ \|x\|_2 = \sqrt{x^{T}x} $ denotes the norm induced by inner product $ \left\langle \cdot \right\rangle $. $ x\otimes y $ denotes their Kronecker product. Given a symmetric positive definite matrix $ \Psi $, denoted as $ \Psi \succ 0 $, denote the $ \Psi $-induced inner product $ \left\langle x,y\right\rangle_{\Psi}=\left\langle \Psi x,y\right\rangle $.
The $ \Psi $-induced norm, $ \| \cdot \|_{\Psi} $, is defined as $ \|x\|_{\Psi} = \sqrt{\left\langle \Psi x,y\right\rangle} $. Denote by $ \|\cdot \| $ any matrix induced norm in the Euclidean space. For $ A\in \mathbb{R}^{w\times m} $, let $ \|A\|=\delta_{max}(A) $ denote the 2-induced matrix norm, where $ \delta_{max}(A) $ is its maximum singular value. If $ A $ is a square matrix, denote $ s_{max}(A) $ and $ s_{min}(A) $ as the maximal eigenvalue and minimum eigenvalue of  $ A $, respectively. Denote $ \mathbf{1}_m = (1, \dots , 1)^{T} \in \mathbb{R}^m  $ and $ \mathbf{0}_m = (0, \dots , 0)^T \in \mathbb{R}^m  $.
Denote  $  diag((A^j)_{j\in \mathbb{N}_{m}^{*} })  $  the block diagonal matrix with $ A^1, \dots, A^m $ on its main diagonal. $ \text{Null} (A) $ and $\text{Range}(A) $ denote the null space and range space of matrix $ A $, respectively.
Denote $ col((x^j_i)_{j\in \mathbb{N}_{m}^{*},i\in \mathbb{N}_{n_{j}}^{*}}):=[x^1_1, \dots ,x^1_{n_1},\dots,x^m_1,\dots, x^m_{n_j}]^T $ as the stacked column vector.
$ I_m $ denotes the identity matrix in $ \mathbb{R}^{m\times m} $. Denote $ \times _{j=1,\ldots,m} \Omega^j$ or $ \prod_{j=1}^{m}\Omega^j $ as the Cartesian product of the sets $ \Omega^j $.
Define the indicator function of $ \Omega^j $ as $ \iota_{\Omega^j}(x)=1$ if $ x\in \Omega^j $ and $ \iota_{\Omega^j}(x)=0 $ if $ x \notin \Omega^j $.
\par
Let $ \mathcal{A}:\mathbb{R}^n\rightarrow 2^{\mathbb{R}^n} $ denotes a set-valued operator \cite{bauschke2011convex}.  The domain of $ \mathcal{A} $ is $ dom (\mathcal{A})=\{x \in \mathbb{R}^n|\mathcal{A}x\neq \emptyset\} $ where $ \emptyset $ stands for the empty set.
The zero set of $ \mathcal{A} $ is $ zer(\mathcal{A})=\{x\in\mathbb{R}^n|\mathbf{0}\in \mathcal{A}x\} $.
The resolvent of the operator $ \mathcal{A} $ is $ J_{\mathcal{A}}=(\text{Id}+\mathcal{A})^{-1} $, where $ \text{Id} $ denotes the identity operator.
An operator $ \mathcal{A} $ is monotone if $ \left\langle \mathcal{A}x-\mathcal{B}y, x-y  \right\rangle  \geq 0 $, and it is maximally monotone if its graph is not strictly contained in the graph of any other
monotone operator, where the graph of $ \mathcal{A} $ is defined as $ gra(\mathcal{A})=\{(x,u)|u \in \mathcal{A}x \} $.
Given a proper, lower semi-continuous, and convex function $ h $, the sub-differential is the operator $\partial h(x): dom(h) \rightarrow 2^{\mathbb{R}^n} $ is $x\longmapsto\{u\in\mathbb{R}^n|h(y) \geq h(x)+\left\langle u,y-x \right\rangle  , \forall y\in dom(h) \}  $.
The proximal operator is defined as $ \text{Prox}_h(v):=\text{argmin}_{u\in dom(h)}\{h(u)+1/2\|u-v\|^2\}=J_{\partial h}(v) $,
and if $ \Psi \succ 0 $, then  $ \text{Prox}^{\Psi}_{h}(v)=J_{\Psi\partial h}(v)  $.
And  $ N_{\Omega} :\mathbb{R}^n\rightarrow 2^{\mathbb{R}^n} $ is a set-valued mapping which denotes the norm cone operator for the set
$ \Omega $, i.e.,
if $ x \in\Omega $, then $ N_{\Omega}=\{ v\in \mathbb{R}^n |  sup_{u\in \Omega } v^T(u-x)\leq 0\} $ ,
otherwise $ N_{\Omega}(x)=\emptyset $.

\section{GAME FORMULATION} \label{section2}
In this section, we introduce the formulation of multi-cluster games.
The full-information and partial-information regimes are introduced in subsection \ref{subfull1} and \ref{subpartial1} respectively.
\subsection{Multi-cluster Games with Full-decision Information} \label{subfull1}
Consider a multi-cluster game consists of $ m $ clusters where each cluster is treated as a  player in a non-cooperative game. Every cluster contains a number of agents, i.e., $n_j $ denotes the number of agents in cluster $ j $ where $ j\in  \mathbb{N}_{m}^{*} $. Denote $  x_i^j\in \mathbb{R}^{q_j}$  the decision of agent $ i $ in cluster $ j $, and $ x^j\in \mathbb{R}^{n_j q_j}  $ a stacked strategy vector of cluster $ j $   piled up by $ x_i^j $, the same as $ \boldsymbol{x}\in \mathbb{R}^q $  piled up by $x^j $. The number of agents in the game is  $ n=\sum_{j=1}^m n_j $ , and the dimensions of $ \boldsymbol{x} $ is the sum of all clusters: $ q=\sum_{j=1}^m n_jq_j$.
\par Each cluster's payoff function is the sum of all agent's payoff functions in that cluster, i.e., cluster $ j $'s payoff function is given by
\[
\theta^{j}\left(x^{j}, \boldsymbol{x}^{-j}\right)=\sum_{i=1}^{n_{j}} \theta_{i}^{j}\left(x^{j}, \boldsymbol{x}^{-j}\right): \mathbb{R}^{q} \rightarrow \mathbb{R},
\]
where $\boldsymbol{x}^{-j}=col((x^j)_{j\in  \mathbb{N}_{m}^{*} \backslash j})$ denotes a stacked strategy vector piled up by all agents' strategies except agents in cluster $ j $,  and $ \theta^j_i:\mathbb{R}^{q} \rightarrow (-\infty,\infty]$ denotes an extended-valued local payoff function of agent $ i $ in cluster $ j $. We usually consider $ \theta^j_i$ can be split into smooth and non-smooth parts, i.e.,
\begin{equation} \label{eq1}
	\theta_{i}^{j}\left(x^{j}, \boldsymbol{x}^{-j}\right)=
	f^{j}_{i}\left(x^{j}, \boldsymbol{x}^{-j}\right)+
	h^{j}_{i}(x^{j}_i).	
\end{equation}
\newtheorem{assumption}{\textbf{Assumption}}
\begin{assumption} \label{assumption1}
	For each $j\in  \mathbb{N}_{m}^{*} $, $ i\in  \mathbb{N}_{n_{j}}^{*}$, and for fixed $ \boldsymbol{x}^{-j} \in  \mathbb{R}^{q-n_j q_j}$, the function
	$ f^j_i(\cdot,\boldsymbol{x}^{-j}) $ in (\ref{eq1}) is convex and continuously differentiable,
	$ h^j_i:\mathbb{R}^{q_j}\rightarrow (-\infty,\infty] $  is lower semi-continuous and convex, and $ dom(h^j_i)=\Omega^j_i$ is a non-empty compact and convex set.
\end{assumption}
\par We assume the first agent in a cluster is a leader node which processes partial information of coupled inequality constraint among clusters. And strategies are assumed to be agreed by all the agents in the same cluster. Hence, the feasibility set $ \mathcal{K}  $ of game's strategy variables is defined as follows
\begin{equation}\label{setK}
	\begin{aligned}
		\mathcal{K}&=\{
		\boldsymbol{x}\in  \Omega | \, \sum_{j=1}^{m}A^{j} x_{1}^{j} \leq \sum_{j=1}^{m}b^{j},\, x_{i}^{j}=x_{l}^{j}\in \mathbb{R}^{q_j} ,\\
		& \qquad \qquad \qquad \qquad  \forall j\in  \mathbb{N}_{m}^{*},\,\forall i,l\in  \mathbb{N}_{n_{j}}^{*}\},	
	\end{aligned}
\end{equation}
where $ A^j \in \mathbb{R}^{w\times q^j}$, $b^j \in \mathbb{R}^w $, $ \Omega=\prod_{j=1}^{m}\Omega^j $, $\Omega^j=\prod_{i=1}^{n_j}\Omega^j_i  $.
 Since each cluster can be treated as a non-cooperative player in the game, then for cluster $ j $, the minimization problem can be written as follows,
\begin{equation}\label{min}
	\begin{aligned}
		\mathop{\text{minimize}}\limits_{x^j}\,\,\, &\theta^j(x^j,\boldsymbol{x}^{-j}),\\
		\text{ s.t. }  \,\,\, &x^j\in \mathcal{K}^j(\boldsymbol{x}^{-j}),
	\end{aligned}
\end{equation}
where $  \mathcal{K}^j(\boldsymbol{x}^{-j}) := \{x^j\in \Omega^j \,|\, (x^j,\boldsymbol{x}^{-j}) \in \mathcal{K} \}$ is the feasible
decision set of agent $ i $ in cluster $ j $. Then a GNE $ \boldsymbol{x}^*=col(x^{j*})_{j\in \mathbb{N}_{m}^{*}} $ of game (\ref{min}) is defined as
\begin{equation} \label{GNE}
	\begin{split}
		&x^{j*} \in \text{argmin }\, \theta^j (x^j,\boldsymbol{x}^{-j*}),	\\
		&\text{s.t.} \quad x^j\in \mathcal{K}^j(\boldsymbol{x}^{-j*}) , \, \forall j\in  \mathbb{N}_{m}^{*}.
	\end{split}
\end{equation}
\begin{assumption}\label{assumption2}
	$ \mathcal{K}$ is non-empty and satisfies Slater's constraint qualification.
\end{assumption}
\par We assume all agents in the same cluster are connected by a graph $ \mathcal{G}^j $ where $ j\in  \mathbb{N}_{m}^{*} $. Moreover, in order to deliver information between clusters, we assume that all leaders are  connected by a graph $ \mathcal{G}_L $, as shown in Fig.1.
Then we make the following assumption.
\begin{assumption}\label{assumption3}
	For all $ j\in \mathbb{N}_{m}^{*} $, $ \mathcal{G}^j $ and $\mathcal{G}_L $	 are undirected and connected.
\end{assumption}
\par We denote  $ \mathcal{G}^j=(\mathcal{N}^j,\mathcal{E}^j)$ is an undirected graph of cluster $ j $, $ \mathcal{N}^j $ is the set of agents, $\mathcal{E}^j\subset \mathcal{N}^j\times \mathcal{N}^j$ is the edge set,
and $ \mathcal{W}^j=[w^j_{i,\xi}]\in \mathbb{R}^{n_j\times n_j} $ is the adjacency matrix such that $ w^j_{i,\xi} = w^j_{\xi,i} >0 $ if $ {j,i}\in\mathcal{E} $ and $ w^j_{i,\xi }=0 $  otherwise \cite{mesbahi2010graph}.
Let  $ d^j_{i,\xi}=\sum_{\xi=1}^{n_j}w^j_{i,\xi} $, then $ Deg^j=diag((d^j_{i,\xi})_{i\in \mathbb{N}_{n_{j}}^{*}})\in\mathbb{R}^{n_j \times n_j} $.
The weighted Laplacian of $ \mathcal{G}^j $ is defined as $ L^j=Deg^j-\mathcal{W}^j $.
If graph $ \mathcal{G}^j $ is undirected and connected, then $ 0 $ is a simple eigenvalue of $ L^j $, $ L^j\boldsymbol{1}_{n_j}=\boldsymbol{0}_{n_j} $, $ \boldsymbol{1}^T_{n_j}L^j=\boldsymbol{0}^T_{n_j} $, and all other eigenvalues are positive.
The eigenvalues of $ L^j $ in ascending order is given by $ 0=s_1(L^j) < s_2(L^j) \leq \ldots \leq  s_{n_j}(L^j)  $.
And we denote $ L=diag((L^j)_{j\in \mathbb{N}^*_m })$.
\par The undirected and connected graph $ \mathcal{G}_L $ is defined similarly. Denote its weighted matrix as $ \mathcal{W}_L =[w^{j,l}_{1,1}]\in \mathbb{R}^{m\times m}$, and Laplacian matrix as $ L^0_m $.
\begin{figure}[htbp]
	\centering
	\includegraphics[width=2.8in]{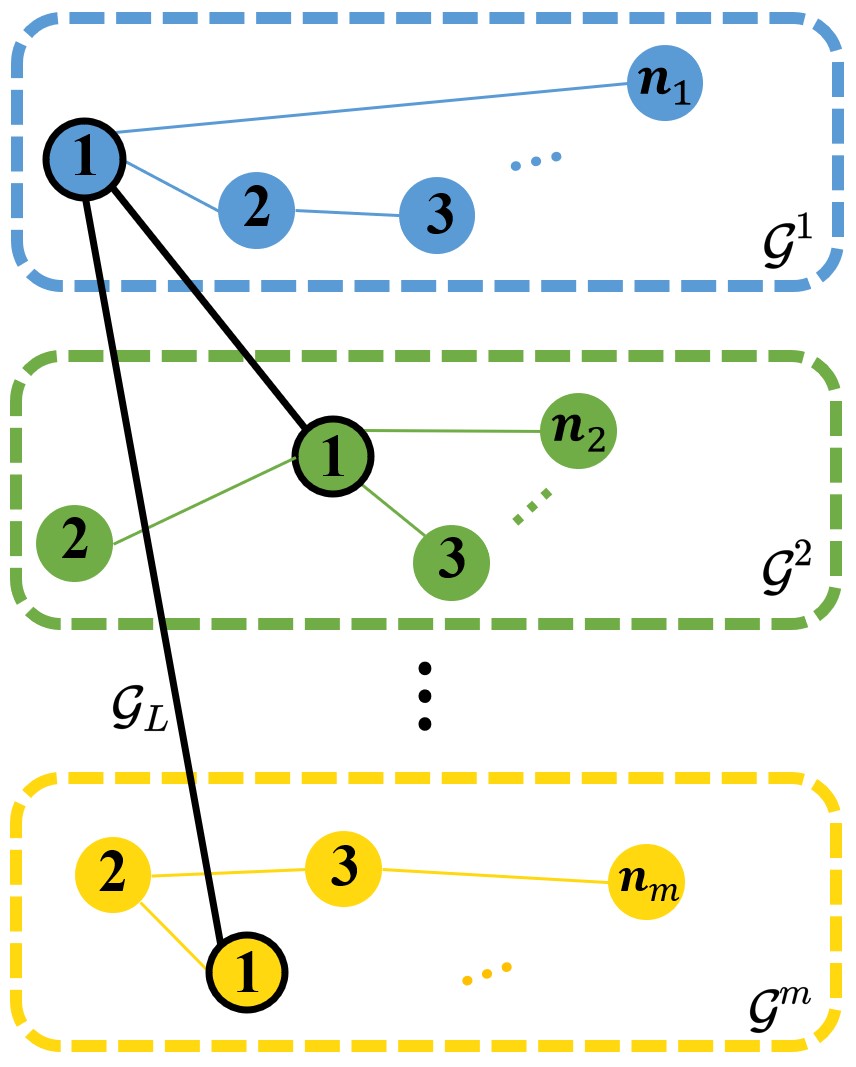}\\
	{Fig.1. The communication topology of the multi-cluster game. }
\end{figure}
\par
We recast the condition $ x^j_i=x^j_l $ in (\ref{setK}) as $ \boldsymbol{L}^jx^j=0 $   \cite{zeng2019},
Then we define a local Lagrangian function for cluster $ j $ with multiplier $\boldsymbol{\lambda}^j\in \mathbb{R}^{n_jq_j} $, $ \mu^j \in \mathbb{R}^w  $ as
\begin{equation}
	\begin{split}
		\mathcal{L}(x^j,\boldsymbol{\lambda}^j,\mu^j;\boldsymbol{x}^{-j} )=\theta^j(x^j,\boldsymbol{x}^{-j})+&
		(\boldsymbol{\lambda}^j)^T \boldsymbol{L}^j x^j\\
		&+(\mu^j)^T (\boldsymbol{A}\boldsymbol{x}-b),
	\end{split}
\end{equation}	
where
$ \boldsymbol{L}^j =L^j \otimes I_{q_j} $, $ b=\sum_{j=1}^{m}b^j$, and
\[
\boldsymbol{A}=\left[A^1,\boldsymbol{0}_{w\times (n_1q_1-q_1)},\ldots,A^m,\boldsymbol{0}_{w\times (n_mq_m-q_m)}\right]\in \mathbb{R}^{w\times q}.
\]
Note that $ \boldsymbol{A}\boldsymbol{x}=\sum_{j=1}^{m}\boldsymbol{A}^jx^j=\sum_{j=1}^{m}A^j x^j_1 $, where $ \boldsymbol{A}^j=\left[A^j,\boldsymbol{0}_{w\times (n_jq_j-q_j)}\right]\in \mathbb{R}^{w\times n_jq_j} $.
\par When $ \boldsymbol{x}^* $ is an GNE of game (\ref{min}), there exist $\boldsymbol{\lambda}^{j*} \in \mathbb{R}^{n_jq_j} $ and $ \mu ^{j*} \in \mathbb{R}^w_+$ such that the following Karush Kuhn Tucker (KKT) \cite{facchinei2007finite} conditions are satisfied:
\begin{equation}
	\left\{\begin{aligned}
		\mathbf{0}_{n_{j} q_{j}} &\in \nabla_{x^{j *}} f^{j}\left(x^{j*}, \boldsymbol{x}^{-j *}\right)
		+\partial_{x^{j *}}h^j(x^j)\\
		&\qquad 	\qquad 	 \qquad  \qquad+\boldsymbol{L}^{j} \boldsymbol{\lambda}^{j *}  +\left(A^{j}\right)^{T} \mu^{j *} \\
		\mathbf{0}_{w}&\in b-\boldsymbol{A} \boldsymbol{x}^{*}+N_{\mathbb{R}_{>0}^{w}}\left(\mu^{j *}\right) \\
		\mathbf{0}_{n_{j} q_{j}}&=\boldsymbol{L}^{j} x^{j *}, \quad j \in \mathbb{N}_{m}^{*},
	\end{aligned}\right.
\end{equation}
where $\nabla_{x^{j}} f^{j}(x^{j}, \boldsymbol{x}^{-j})=\sum_{i=1}^{n_j}\nabla_{x^j}f^j_i (x^j, \boldsymbol{x}^{-j})$, 
and $\partial_{x^j}h^j(x^j)=\sum_{i=1}^{n_j}\partial_{x^{j}}h^j_i(x^j_i) $.  
\par  In this work, we seek GNE with the same Lagrangian multiplier $ \mu $, i.e., $ \mu^1=\mu^2=\dots=\mu^{m-1}=\mu^m=\mu $ for all non-cooperative participants, called variational GNE (v-GNE) which is a subclass of GNE, and $ \boldsymbol{x}^*\in \mathcal{K} $ is a solution of the variational inequality GVI$ (\Theta,\mathcal{K}) $:
\begin{equation}\label{VI}
	\exists\theta_{\boldsymbol{x}^*} \in \Theta(\boldsymbol{x}^*) \,\,\left\langle \theta_{\boldsymbol{x}^*},\boldsymbol{x}-\boldsymbol{x}^{*}\right\rangle \geq 0, \quad \forall \boldsymbol{x} \in \mathcal{K},	
\end{equation}
where $ \Theta(\boldsymbol{x}^*)=F (\boldsymbol{x}^{*}) + H(\boldsymbol{x}^*) $,  and $ F  $ is a mapping of the game which is given by
\begin{equation}\label{psedugradient}
	F(\boldsymbol{x})=col((\nabla_{x^{j}}f^j(x^j,\boldsymbol{x}^{-j}))_{j\in  \mathbb{N}_{m}^{*}}),
\end{equation}
and $ H(\boldsymbol{x})=\partial_{x^{1}}h^1(x^1)\times \dots \times \partial_{x^{m}}h^m(x^m) $.
\par Before  discussing  the existence and uniqueness of a v-GNE, we make the following assumption.
\begin{assumption}\label{assumption4strongmonotone}
	$ F $ is strongly monotone and Lipschitz continuous: for any $ \boldsymbol{x},\boldsymbol{x}'$, there existed $ \eta>0 $
	$ \left\langle F(\boldsymbol{x}) - F(\boldsymbol{x}'), \boldsymbol{x} - \boldsymbol{x}'\right\rangle \geq \eta \| \boldsymbol{x} - \boldsymbol{x}'\|^2$, and there exist $ \kappa_0 >0 $
	such that $ \|F(\boldsymbol{x}) - F(\boldsymbol{x}')\| \leq \kappa_0 \|\boldsymbol{x} - \boldsymbol{x}'\|   $.
\end{assumption}
\par Let Assumption \ref{assumption1}, \ref{assumption2}, \ref{assumption4strongmonotone} be satisfied.
Note that $ H $ is maximally monotone since it is  constituted by sub-differential of lower semi-continuous convex function $ h^j_i $, for all $ j\in \mathbb{N}_{m}^{*}  $, $ i\in  \mathbb{N}_{n_{j}}^{*} $.
Then $ \Theta =F+H $ is monotone since the direct sum of maximally monotone is monotone (Prop. 20.23, \cite{bauschke2011convex}).
Then it follows by (Prop.12.11, \cite{facchi201012}) that since $\Theta  $
is  monotone then there exists  a solution to GVI$ (\Theta, \mathcal{K}) $. And it also guarantees the existence of a v-GNE.
\begin{assumption} 	\label{solutionGVI}
	The  solution set of GVI$ (\Theta, \mathcal{K}) $ is nonempty.
 \end{assumption}
\par More sufficient conditions for the existence of solutions to monotone GVI can be found in \cite{facchi201012}, the compactness of $ \Omega^j, j\in \mathbb{N}^*_m $ also ensures the Assumption \ref{solutionGVI}.
\par If $ \boldsymbol{x}^* $ is the solution of GVI$ (\Theta,\mathcal{K}) $, then there exist $ \boldsymbol{\lambda}^*\in \mathbb{R}^q $ and $ \mu^* \in \mathbb{R}^w $ such that the following KKT conditions are satisfied:
\begin{equation}\label{vKKT}
	\left\{\begin{array}{c}
		\mathbf{0}_{q} \in F\left(\boldsymbol{x}^{*}\right)+H\left(\boldsymbol{x}^{*}\right)+\boldsymbol{L} \boldsymbol{\lambda}^{*}+\boldsymbol{A}^{T} \mu^{*}	 \\
		\mathbf{0}_{w} \in b-\boldsymbol{A} \boldsymbol{x}^{*}+N_{\mathbb{R}_{>0}^{w}}\left(\mu^{*}\right) \\
		\mathbf{0}_{q}=\boldsymbol{L} \boldsymbol{x}^{*}, \quad j \in \mathbb{N}_{m}^{*} ; k, l \in \mathbb{N}_{n_{j}}^{*},
	\end{array}\right.
\end{equation}
where $ \boldsymbol{L}=diag((\boldsymbol{L}^j)_{j\in \mathbb{N}_{m}^{*}}) $,
  $ \boldsymbol{\lambda}=col((\boldsymbol{\lambda}^j)_{j\in \mathbb{N}_{m}^{*}} )$.
\newtheorem{lemma}{\textbf{Lemma}}
\begin{lemma}	\label{vGNEisGNE}
	Let Assumptions \ref{assumption1}, \ref{assumption2}, \ref{assumption4strongmonotone} hold. Then $ \boldsymbol{x}^*\in\mathbb{R}^q $ is a v-GNE of the game (\ref{min}) if and only if there exists $\boldsymbol{\lambda}^*\in \mathbb{R}^q $ and $ \mu^* \in \mathbb{R}^w $ such that $ (\boldsymbol{x}^*,\boldsymbol{\lambda}^*,\mu^*) $ satisfied the KKT conditions in (\ref{vKKT}).
\end{lemma}
\par \textbf{Proof}:
	Assumptions
	\ref{assumption1},
	\ref{assumption2}, \ref{assumption4strongmonotone} guarantees the existence of a solution to
	GVI$ (\Theta,\mathcal{K}) $,
	by (Prop.12.11 \cite{facchi201012},
	Prop.2.3 \cite{hu2015multi}). For any  $(\boldsymbol{x}^*,\boldsymbol{\lambda}^*,\mu^*) $ satisfied the KKT conditions in (\ref{vKKT}), the same $ \boldsymbol{x}^* $ is the solution of GVI$ (\Theta,\mathcal{K}) (\ref{VI}) $ by
	( Prop.3.3 \cite{belgioioso2020equilibrium},
	Th.3.1\cite{auslender2000lagrangian}).
	And every solution $ \boldsymbol{x}^* $ of GVI$ (\Theta,\mathcal{K}) (\ref{VI}) $ is a GNE of game (\ref{min}) by
	(Prop.12.4 \cite{facchi201012},
	Th.3.9 \cite{facchi2010GNE}).
	Then the same $ \boldsymbol{x}^* $ in (\ref{vKKT}) is a v-GNE of game (\ref{min}).
	\hfill $\blacksquare$
\subsection{Multi-cluster Games with Partial-decision Information} \label{subpartial1}
\par In the following, we consider a more realistic scenario that the agents cannot have full decision information on others outside the cluster $ \boldsymbol{x}^{-j} $, called partial-decision information scenario. Under this setting, agents have to estimate the strategies of all other agents outside cluster by relying on the information exchanged with some neighbors over communication networks $ \mathcal{G}^j,\forall{j\in  \mathbb{N}_{m}^{*}} $ and
$ \mathcal{G}_L $.
\par  In cluster $ j $, $ \hat{x}^{j,j'}_{i,i'}\in \mathbb{R}^{q_{j'}} $ denotes the estimation of agent $ i $  to agent $ i' $ in cluster $ j' $.
Then the  estimation vector of agent $ i $ in cluster $ j $ to all other agents is given by
\[
\hat{\boldsymbol{x}}^{j}_{i} =col((\hat{x}^{j,j'}_{i,i'})_{j'\in  \mathbb{N}_{m}^{*} ,i'\in \mathbb{N}_{n_{j'}}^{*}})\in \mathbb{R}^{q}.	
\]
Agents in the same cluster  are cooperators, so we assume they can obtain information of others by a conference graph or a virtual center,
then for agent $ i  $ in cluster $ j $, the estimations of cluster $ j $ are the real value, i.e., $\hat{x}^{j,j}_{i,i'}=x^j_{i'}$.
And  the strategy profiles can be stacked up
$ \hat{\boldsymbol{x}}^{j}=col((\hat{\boldsymbol{x}}^{j}_{i})_{i\in{ \mathbb{N}_{n_{j}}^{*}}})$,
$ \hat{\boldsymbol{x}}=col((\hat{\boldsymbol{x}}^j)_{j\in \mathbb{N}_{m}^{*}})$.
Moreover, the  estimation vector of agent $ i $ in cluster $ j $ to all agent's except cluster $ j $ denotes as
$ \hat{\boldsymbol{x}}^{j,-j}_i =col((\hat{x} ^{j,j'}_{i,i'} )_{j'\in  \mathbb{N}_{m}^{*}\backslash j,\,i'\in  \mathbb{N}_{n_{j'}}^{*}}) \in \mathbb{R}^{q-n_jq_j}$.
In the same way, the two stacked vectors are denoted as
$ \hat{\boldsymbol{x}}^{j,-j}=col((\hat{\boldsymbol{x}}^{j,-j}_i)_{i\in  \mathbb{N}_{n_{j}}^{*}}) $,  $\hat{\boldsymbol{x}}^{-j}=col((\hat{\boldsymbol{x}}^{j,-j})_{j\in \mathbb{N}_{m}^{*}} ) $.
Next, we give special matrices $ S^j_i $ and $ R^j_i $ to extract values from $\hat{\boldsymbol{x}}^j_i  $, and they are defined as
\begin{equation}\label{RS}
	\begin{aligned}
		&{\small S_{i}^{j}}=\left[\begin{smallmatrix}
			\boldsymbol{I}_{{  \left(n_{<j} q_{<j}\right)}}& \mathbf{0}_{{  \left(n_{<j} q_{<j}\right) \times\left(n_{j} q_{j}\right)}}& \mathbf{0}_{{  \left(n_{<j} q_{<j}\right) \times\left(n_{>j} q_{>j}\right)}} \\
			\mathbf{0}_{{  \left(n_{>j} q_{>j}\right) \times\left(n_{<j} q_{<j}\right)}}& \mathbf{0}_{{  \left(n_{>j} q_{>j}\right) \times\left(n_{j} q_{j}\right)}}& \boldsymbol{I}_{{ \left(n_{>j} q_{>j}\right)}}
		\end{smallmatrix}\right]\\
			&{\small R_{i}^{j}}=\left[\begin{smallmatrix}
		\boldsymbol{0}_{q_{j} \times\left(n_{<j} q_{<j}+(i-1) q_{j}\right)} \,\, \boldsymbol{I}_{q_{j}} \,\, \boldsymbol{0}_{q_{j} \times\left(\left(n_{j}-i\right) q_{j}+n_{>j} q_{>j}\right)}
	\end{smallmatrix}\right],		
	\end{aligned}
\end{equation}
where $ n_{<j}q_{<j}=\sum_{j'<j, j' \in \mathbb{N}_{m}^{*}} n_{j'} q_{j'} $ and $  n_{>j}q_{>j} = \sum_ {j'>j, j' \in \mathbb{N}_{m}^{*}} n_{j'} q_{j'}  $.
Note that $ R_i^j $ selects the  local strategy vector of agent $ i $ in cluster $ j $ from its estimation of all agents in the game, i.e., $ R_{i}^{j} \hat{\boldsymbol{x}}_{i}^{j}=x_{i}^{j} $, while  $ S_i^j $ removes the whole cluster $ j $'s strategy vector from $ \hat{\boldsymbol{x}}_{i}^{j} $, i.e., $ S_{i}^{j} \hat{\boldsymbol{x}}_{i}^{j}=\hat{\boldsymbol{x}}_{i}^{j,-j} $. And we denote
$ R^j=diag((R^j_i)_{i \in  \mathbb{N}_{n_{j}}^{*}})$, $ R=diag((R^j)_{j \in  \mathbb{N}_{m}^{*}})$, $ S^j=diag((S^j_i)_{i \in  \mathbb{N}_{n_{j}}^{*}}) $,
$ S^j=diag((S^j_i)_{j \in  \mathbb{N}_{m}^{*}})$, then it's easy to  derive that
$ R^j\hat{\boldsymbol{x}}^j=\boldsymbol{x}^j $, $ R\hat{\boldsymbol{x}}=\boldsymbol{x} $, $ S^j\hat{\boldsymbol{x}}^j=\hat{\boldsymbol{x}}^{j,-j}$, $ S\hat{\boldsymbol{x}}=\hat{\boldsymbol{x}}^{-j} $.
\par  For partial-decision scenario, the minimization problem of cluster $ j $ in (\ref{min}) is rewritten as:
\begin{equation}\label{min2}
	\begin{array}{ll}
		\text{minimize} & \sum_{i=1}^{n_{j}} f_{i}^{j}(x^{j}, \hat{\boldsymbol{x}}_{i}^{j,-j})+\sum_{i=1}^{n_{j}} h_{i}^{j}(x^{j}_i) \\
		\text { s.t. } & \boldsymbol{L}^{j} R^{j} \hat{\boldsymbol{x}}^{j}=\mathbf{0}_{n_{j} q_{j}} \\
		& \boldsymbol{A}^{j} R^{j} \hat{\boldsymbol{x}}^{j} \leq b-\sum_{j \in \mathbb{N}_{m}^{*} \backslash\{j\}} \boldsymbol{A}^{h} R^{h} \hat{\boldsymbol{x}}^{h},
	\end{array}
\end{equation}
\par The Lagrangian function of partial-decision scenario is defined as
\begin{equation}
	\begin{split}
		\mathcal{L}^{j}  (R^{j} \hat{\boldsymbol{x}}^{j},   \hat{\boldsymbol{x}}^{j,-j}  )=&\theta^{j}  (R^{j}   \hat{\boldsymbol{x}}^{j}, \hat{\boldsymbol{x}}^{j,-j}  )+  (\boldsymbol{\lambda}^{j}  )^{T} \boldsymbol{L}^{j} R^{j} \hat{\boldsymbol{x}}^{j}\\
		&\qquad +(\mu^{j})^T  (\boldsymbol{A} R \hat{\boldsymbol{x}}-b ).
	\end{split}
\end{equation}
\par When $\hat{\boldsymbol{x}}^*$ is a v-GNE of partial-decision scenario in the game (\ref{min2}), then the following KKT  conditions \cite{facchinei2007finite} must be satisfied:
\begin{equation} \label{pKKT}
	\left\{\begin{array}{c}
		\mathbf{0}_{n q} \in R^{T} \boldsymbol{F}\left(  \hat{\boldsymbol{x}}^{*}\right)+R^{T} H\left(R   \hat{\boldsymbol{x}}^{*}\right)+R^{T} \boldsymbol{L} \boldsymbol{\lambda}^{*}+R^{T} \boldsymbol{\Lambda}^{\mathrm{T}} \boldsymbol{\mu}^{*} \\
		\mathbf{0}_{w}\in b-\boldsymbol{A} R   \hat{\boldsymbol{x}}^{*}+N_{\mathbb{R}_{>0}^{w}}\left(\mu^{*}\right) \\
		\mathbf{0}_{n q}=\boldsymbol{L} R   \hat{\boldsymbol{x}}^{*},
	\end{array}\right.
\end{equation}
where $  \boldsymbol{F}\left( \hat{\boldsymbol{x}}\right) = col((\nabla_{x^j}f^j(x^j,\hat{\boldsymbol{x}}^{j,-j}))_{j\in  \mathbb{N}_{m}^{*}}) $
  is the extended mapping of $ F(\boldsymbol{x}) $  as \cite{pavel2019distributed} defined,
  and $ \boldsymbol{\mu}=\mathbf{1}_m \otimes \mu \in \mathbb{R}^{wm} $,
  $\boldsymbol{\Lambda}=diag((\boldsymbol{A}^{j})_{j \in \mathbb{N}_{m}^{*}}) \in \mathbb{R}^{wm \times q} $.

\begin{assumption}\label{assumption5extendgradient}
  	The extended pseudo-gradient $ \boldsymbol{F} $, as defined in (\ref{pKKT}), is Lipschitz continuous, i.e.,  there exists $ \kappa >0$ such that for any $ \hat{\boldsymbol{x}} $ and $ \hat{\boldsymbol{x}}' $,
  	$ \|\boldsymbol{F}(\hat{\boldsymbol{x}})- \boldsymbol{F}(\hat{\boldsymbol{x}}')\|\leq \kappa \|\hat{\boldsymbol{x}}-\hat{\boldsymbol{x}}'\| $.
 \end{assumption}
\par Note that a v-GNE of (\ref{min2}) is given by $ \hat{\boldsymbol{x}}^* =\boldsymbol{1}_n \otimes \boldsymbol{x}^*$, while the same $ \boldsymbol{x}^* $ is the solution of (\ref{VI}), and the v-GNE of (\ref{vKKT}). We show the details in convergence analysis section.
\section{DISTRIBUTED ALGORITHM} \label{section3}
\par In this section, we present a distributed algorithm based on Forward-Backward-Forward approach.
\par In the scenario of partial-decision,  agent $ i $ in cluster $ j $ have to estimate  other agents' strategies. However, it only estimates strategies  outside cluster $ j $, i.e., $ \hat{\boldsymbol{x}}^{j,-j}_i \in \mathbb{R}^{q-n_jq_j}$ , since it already processes its local strategy profile $ x^j \in \mathbb{R}^{n_jq_j}$.
Moreover, $ \lambda^j_i \in \mathbb{R}^{q_j}  $ and $ \mu^j \in \mathbb{R}^{w} $ are dual multipliers for the estimation of components of vector  $ \boldsymbol{\lambda}^* $ and $ \boldsymbol{\mu}^* $ respectively. And $ z^j\in \mathbb{R}^w $ denote a local auxiliary variable used for the coordination needed to satisfy the coupling constraint and to reach consensus of the dual variable $ \mu^j $. Moreover, $\lambda^j_i  $ is a variable used to make all agent's strategies to come consensus while these agents are in the same cluster.
\par In Algorithm \ref{algorithm}, we split iteration $ k\rightarrow k+1 $ into two procedures: $ k\rightarrow k' $ and $ k'\rightarrow k+1 $.
\par \textbf{In procedure $  k\rightarrow k' $},  if $ i=1 $ (leader), then agent obtains information of real strategies  and gradient information of other agents in the same cluster $  j\in  \mathbb{N}_{m}^{*}  $ from virtual center to calculate $ v^j_i  $ and $ \tilde{v}^j_i $, where $ v^j_i :=\nabla_{x^j_i}f^j_i(x^j,\hat{\boldsymbol{x}}^{j,-j}_i) $ and $ \tilde{v}^j_i:=\textstyle\sum_{\xi=1}^{n_j}\nabla_{x_{i}^{j}} f_{\xi}^{j}(x^{j}  , \hat{\boldsymbol{x}}_{i}^{j,-j}) $.
Moreover, agent $ i$ exchanges local estimation information of other clusters' agents $ \hat{\boldsymbol{x}}^{j,-j}_i $   and multiplier information $ \lambda^j_i $
via $ \mathcal{G}^j=\{ \mathcal{N}^{j},\mathcal{E}^j\}$.
And $ (\xi,i)\in \mathcal{E}^j $ hold if agent $ i$ in cluster $ j $ receives $ \{ \hat{\boldsymbol{x}}^{j,-j}_{\xi},\lambda^j_{\xi}\} $  from the neighboring agent $ \xi $ in cluster $ j $, where   $\xi \in \mathcal{N}^j_{(i) }:=\{\xi\,|\,(\xi,i)\in \mathcal{E}^j\}  $. The updating procedure of $ x^j_i $, $ \hat{\boldsymbol{x}}^{j,-j}_i  $, $ \lambda^j_i$ is given by:
	\begin{align}
		&x_{i}^{j}[k']=\operatorname{Prox}_{h_{i}^{j}}^{\rho_{i}^{j}}
		(x^j_{i}[k]-\rho^j_i(\tilde{v}^j_i[k]+\iota_{\{1\}}(i)(A^{j})^{T} \mu^{j}[k] \nonumber \\
		& +c \sum_{{l \in \mathcal{N}_{L}^{(j)}}} w^{j,l}_{i,1}(x_{1}^{j}[k]-\hat{x}_{1, 1}^{l, j}[k])
		+\sum_{\xi \in \mathcal{N}^{j}_{(i)}} w_{i,\xi}^{j}(\lambda_{i}^{j}[k]-\lambda_{\xi}^{j}[k]))),\label{dis_x}	
	\end{align}
\begin{align}
&\hat{\boldsymbol{x}}_{i}^{j,-j}[k']=\hat{\boldsymbol{x}}_{i}^{j,-j}[k]
-c \rho_{i}^{j} \sum_{l \in \mathcal{N}_{L}^{(j)}} w_{i,1}^{j, l}(\hat{\boldsymbol{x}}_{1}^{j,-j}[k]-\hat{\boldsymbol{x}}_{1}^{l,-j}[k])  \nonumber \\
&-c \rho_{i}^{j}  \sum_{\xi \in \mathcal{N}^{j}_{(i)}}w_{i, \xi}^{j}(\hat{\boldsymbol{x}}_{i}^{j,-j}[k]-\hat{\boldsymbol{x}}_{\xi}^{j,-j}[k]),\label{dis_xneg}
\end{align}
\begin{align}
 &\lambda_{i}^{j}[k']=\lambda_{i}^{j}[k]+\tau_{i}^{j} \sum_{\xi \in \mathcal{N}^{j}_{(i)}} w_{i,\xi}^{j}(x_{i}^{j}[k]-x_{\xi}^{j}[k]),\label{dis_lambda}
\end{align}		
 where $ x_{i}^{j}[k] $, $ \hat{\boldsymbol{x}}_{i}^{j,-j}[k]  $, $ \lambda_{i}^{j}[k] $ denote $ x_{i}^{j} $, $ \hat{\boldsymbol{x}}_{i}^{j,-j}  $, $ \lambda_{i}^{j} $ at iteration $ k $, and we use $ x_{i}^{j}[ k'] $, $ \hat{\boldsymbol{x}}_{i}^{j,-j}[ k']  $, $ \lambda_{i}^{j}[ k'] $ denote iteration result after iteration $ k $, and $ \rho^j_i $, $ \tau^j_i $, $ \sigma^j $, $ \upsilon^j $ are fixed constant step-sizes of  agent $ i $ in cluster $ j $, and $ c $ is a positive parameter, i.e., $ c>0 $.
 If  $ i \neq 1 $, then $ w^{j,l}_{i,1} =0$. And $ \iota_{\{1\}}(i) $ is an indicator function,
if $ i \neq 1  $, then $ \iota_{\{1\}}(i)=0 $.
\par For leader agent in cluster $ j $, it also exchanges its estimation information  $ \hat{x}^{j,l}_{1,1}  $, $ \hat{\boldsymbol{x}}^{j,-j}_{1} $, multiplier information $ z^j $ and auxiliary information $ \mu^j $ via leader-cluster graph  $ \mathcal{G}_L=\{ \mathcal{N}_{L},\mathcal{E}_L \}$.
And $  (l,j)\in \mathcal{E}_L $ hold if it receives
$ \{\hat{x}^{l,j}_{ 1,1},\hat{\boldsymbol{x}}^{l,-l}_1,z^l,\mu^l\}$ from leader agent in cluster $ l $, where $ l \in \mathcal{N}_L^{(j)}:=\{l\,|\,(l,j)\in \mathcal{E}_L\} $. The updating procedure of $ z^j $ and $ \mu^j $ are given as follows:
\begin{align}
   &z^{j}[k']=z^{j}[k]-\sigma^{j} \sum_{l \in \mathcal{N}_{L}^{(j)}}w^{j,l}_{1,1}(\mu^{j}[k]-\mu^{l}[k]), \label{dis_z}\\
   & \mu^{j}[k']=P_{\mathbb{R}_{+}^{w}}(\mu^{j}[k]+v^{j}(\sum_{l \in \mathcal{N}_{L}^{(j)}}w^{j,l}_{1,1}(z^{j}[k]-z^{l}[k])\nonumber\\
   &\qquad\qquad\qquad\qquad+A^{j} x^{j}_1[k]-b^{j})),\label{dis_mu}
\end{align}
where $z^{j}[k] $, $  \mu^{j}[k] $ denote  $z^{j} $, $  \mu^{j} $ at iteration $ k $,  and we use $z^{j}[ k'] $, $  \mu^{j}[ k'] $ denote first iteration result after iteration $ k $.
\par If $ i\neq 1 $ (not leader), then agent obtains other agents' strategies and gradient information in the same cluster $ j\in  \mathbb{N}_{m}^{*}  $ from virtual center, exchanges
$ \{\hat{\boldsymbol{x}}^{j,-j}_i,\lambda^j_i \} $ and receives $ \{\hat{\boldsymbol{x}}^{j,-j}_{\xi},\lambda^j_{\xi}\} $ via graph
$ \mathcal{G}^j$.
Then $ x^j_i[k']  $ is updated by (\ref{dis_x}) while $ \iota_{\{1\}}(i)=0 $ and $ w^{j,l}_{i,1} =0$; $ \hat{\boldsymbol{x}}^{j,-j}_i[k'] $ is updated by (\ref{dis_xneg}) while $ w^{j,l}_{i,1} =0$; $ \lambda^j_i[k'] $ is updated by (\ref{dis_lambda}). Then the procedure $ k\rightarrow k' $ ends.
\par \textbf{In procedure $ k'\rightarrow k+1 $}, for the ease of notation, we denote
\begin{equation}\label{procedure2}
	(x_{1}^{j}-\hat{x}_{ 1,1}^{l, j}){[k-k']} := (x_{1}^{j}[k]-\hat{x}_{ 1,1}^{l, j}[k])-(x_{1}^{j}[k^{\prime}]-\hat{x}_{ 1,1}^{l, j}[k^{\prime}]),
\end{equation}
and $  \mu^{j}[k-k'] $, $ \tilde{v}^j_i{[k-k']} $,
$(\hat{\boldsymbol{x}}_{i}^{j,-j}-\hat{\boldsymbol{x}}_{\xi}^{j,-j}){[k-k']}  $,
$ (\hat{\boldsymbol{x}}_{1}^{j,-j}-\hat{\boldsymbol{x}}_{1}^{l,-j}){[k-k']} $,
$ (x_{i}^{j}-x_{\xi}^{j}){\left[k-k^{\prime}\right]} $,
$ (\mu^{j}-\mu^{l}){[k-k^{\prime}]} $,
$ (z^{j}-z^{l}){[k-k^{\prime}]} $,
$ (x_{1}^{j}){[k-k^{\prime}]} $ are defined similarly.
\par Then if $ i=1 $ (leader), then agent obtains the newest information of real strategies $ x^j_{i'} $ and gradient information $ v^j_{i'} $ of other agents in the same cluster  from virtual center,  $ \hat{\boldsymbol{x}}^{j,-j}_{\xi},\lambda^j_{\xi} $ from neighbors of inner-cluster  graph,  $ \hat{x}^{l,j}_{ 1,1} $,
$ \hat{\boldsymbol{x}}^{j,-j}_i $, $ z^l, \mu^l $ from neighbors of leader-cluster graph at iteration $ k' $. The second updating procedure of $ x^j_i $, $ \hat{\boldsymbol{x}}^{j,-j}_i  $, $ \lambda^j_i$, $ z^j_i $, $ \mu^j_i $ is given as:
  	\begin{align}
  		 &x_{i}^{j}[k+1]= x_{i}^{j}[k']+c \rho_{i}^{j} \sum_{l \in \mathcal{N}_{L}^{(j)}} w_{i,1}^{j, l}(x_{1}^{j}-\hat{x}_{ 1,1}^{l, j}){[k-k']} \nonumber \\
  		&  \qquad +\rho_{i}^{j}( \sum_{\xi \in \mathcal{N}^{j}_{(i)}} w_{i,\xi}^{j}(\lambda_{i}^{j}-\lambda_{\xi}^{j}){[k-k']}+\tilde{v}^j_i{[k-k']} \nonumber\\
  		& \qquad +\iota_{\{1\}}(i) (A^{j})^{T}\mu^{j}[k-k']), \label{dis_sec_x}
     \end{align}
  	\begin{align}
  		&\hat{\boldsymbol{x}}_{i}^{j,-j}[k+1]=c \rho_{i}^{j} (w_{i, \xi}^{j}\sum_{\xi \in \mathcal{N}^{j}_{(i)}}(\hat{\boldsymbol{x}}_{i}^{j,-j}-\hat{\boldsymbol{x}}_{\xi}^{j,-j}){[k-k']} \nonumber\\
  		& \qquad  +w_{i, 1}^{j, l} \sum_{l \in \mathcal{N}_{L}^{(j)}}(\hat{\boldsymbol{x}}_{1}^{j,-j}-\hat{\boldsymbol{x}}_{1}^{l,-j}){[k-k']})+\hat{\boldsymbol{x}}_{i}^{j,-j}[k'],\label{dis_sec_xneg}
  	\end{align}
  \begin{align}
  		&\lambda_{i}^{j}[k+1]=\lambda_{i}^{j}[k']-\tau_{i}^{j}(\sum_{\xi \in \mathcal{N}^{j}_{(i)}} w_{i, \xi}^{j}(x_{i}^{j}-x_{\xi}^{j}){\left[k-k^{\prime}\right]}),\label{dis_sec_lambda}\\
  		& z^{j}[k+1]=z^{j}[k']+\sigma^{j}(\sum_{l \in \mathcal{N}_{L}^{(j)}} w_{1, 1}^{j, l}(\mu^{j}-\mu^{l}){[k-k^{\prime}]}), \label{dis_sec_z}
  	\end{align}
  \begin{align}
  		& \mu^{j}[k+1]=\mu^{j}[k^{\prime}]-v^{j}(\sum_{l \in \mathcal{N}_{L}^{(j)}} w^{j,l}_{1,1}(z^{j}-z^{l}){[k-k^{\prime}]}\nonumber\\
  		&\qquad   - A^{j}x_{1}^{j}{[k-k^{\prime}]}), \label{dis_sec_mu}
  	\end{align}
  where $ x_{i}^{j}[k+1] $, $ \hat{\boldsymbol{x}}_{i}^{j,-j}[k+1]  $, $ \lambda_{i}^{j}[k+1] $, $ z^j_i[k+1] $, $ \mu^j_i[k+1] $ denote $ x_{i}^{j} $, $ \hat{\boldsymbol{x}}_{i}^{j,-j}  $, $ \lambda_{i}^{j} $ at iteration $ k+1 $.
  \par If $ i\neq 1 $ (not leader), then agent obtains $ x^j_{i'} $, $ v^j_{i'} $ from virtual center, $ \hat{\boldsymbol{x}}^{j,-j}_i $, $ \lambda^j_i $ from neighbors of inner-cluster graph at iteration $ k' $. $ x^j_i[k+1] $ is updated by (\ref{dis_sec_x}) while $ \iota_{\{1\}}(i)=0 $ and $w^{j,l}_{i,1}=0 $; $ \hat{\boldsymbol{x}}^{j,-j}_i[k+1] $ is updated by (\ref{dis_sec_xneg}) while $ w^{j,l}_{i,1}=0 $; $ \lambda^j_i[k+1]$ is updated by (\ref{dis_sec_lambda}). Then the procedure $ k'\rightarrow k+1 $ ends.
\begin{algorithm}[!htbp]
	\caption{ Distributed Multi-cluster Game with Partial-decision Information Algorithm (DMGPA) }\label{algorithm}	
	\textbf{Initialization}: $ x_{i}^{j}[0]\in \Omega^j_i \subset  \mathbb{R}^{q_j} $, $ \hat{\boldsymbol{x}}_{i}^{j,-j}[0] \in \mathbb{R}^{q-n_jq_j} $, $ \lambda_{i}^{j}[0]\in \mathbb{R}^{q_j} $, $z^{j}[0] \in \mathbb{R}^w  $, $  \mu^{j}[0] \in \mathbb{R}^{w} $.
	
	\textbf{Iteration} \textbf{$ k\rightarrow k' $}:	
	
	(1) leader agent in cluster $ j $ receives $\{x^j_{i'}[k],  v^j_{i'}[k] \}_{i'\in \mathbb{N}_{n_{j}}^{*}\backslash i}  $ from virtual center,
	$ \{\hat{x}^{l,j}_{ 1,1}[k],\hat{\boldsymbol{x}}^{l,-l}_1[k],z^l[k],\mu^l[k]\}$ from neighbors in leader-cluster graph,
	$ \{ \hat{\boldsymbol{x}}^{j,-j}_{\xi}[k],\lambda^j_{\xi}[k]\} $ from neighbors in inner-cluster graph, then:
	
\textit{	\textbf{updates} $ x^j_i[k'] $ by (\ref{dis_x}); $ \hat{\boldsymbol{x}}^{j,-j}_{i}[k'] $ by (\ref{dis_xneg}); $ \lambda^j_i[k'] $ by (\ref{dis_lambda});
	$ \mu^j[k']$ by (\ref{dis_z}); $ z^j[k'] $  by (\ref{dis_mu});}
	
	(2) non-leader  in cluster $ j $ receives $\{x^j_{i'}[k],  v^j_{i'}[k] \}_{i'\in \mathbb{N}_{n_{j}}^{*}\backslash i}  $ from virtual center,
	$ \{ \hat{\boldsymbol{x}}^{j,-j}_{\xi}[k],\lambda^j_{\xi}[k]\} $ from neighbors in inner-cluster graph, then:
	
\textit{    \textbf{updates} $ x^j_i[k'] $, $ \hat{\boldsymbol{x}}^{j,-j}_{i}[k'] $ by (\ref{dis_x}) and (\ref{dis_xneg}) while  $ \iota_{\{1\}}(i)=0 $ and $ w^{j,l}_{i,1}=0 $; $ \lambda^j_i[k'] $ by (\ref{dis_lambda});}

\textbf{Iteration} \textbf{
	$ k' \rightarrow k+1 $}:

(1) leader agent in cluster $ j $ receives $\{x^j_{i'}[k'],  v^j_{i'}[k'] \}_{i'\in \mathbb{N}_{n_{j}}^{*}\backslash i}  $ from virtual center,
$ \{\hat{x}^{l,j}_{ 1,1}[k'],\hat{\boldsymbol{x}}^{l,-l}_1[k'],z^l[k'],\mu^l[k']\}$ from neighbors in leader-cluster graph,
$ \{ \hat{\boldsymbol{x}}^{j,-j}_{\xi}[k'],\lambda^j_{\xi}[k']\} $ from neighbors in inner-cluster graph, then:

\textit{	\textbf{updates} $ x^j_i[k+1] $ by (\ref{dis_sec_x}); $ \hat{\boldsymbol{x}}^{j,-j}_{i}[k+1] $ by (\ref{dis_sec_xneg}); $ \lambda^j_i[k+1] $ by (\ref{dis_sec_lambda});
	$ \mu^j[k+1]$ by (\ref{dis_sec_mu}); $ z^j[k+1] $  by (\ref{dis_sec_z});}

(2) non-leader  in cluster $ j $ receives $\{x^j_{i'}[k'],  v^j_{i'}[k'] \}_{i'\in \mathbb{N}_{n_{j}}^{*}\backslash i}  $ from virtual center,
$ \{ \hat{\boldsymbol{x}}^{j,-j}_{\xi}[k'],\lambda^j_{\xi}[k'] \}$ from neighbors in inner-cluster graph, then:

\textit{    \textbf{updates} $ x^j_i[k+1] $, $ \hat{\boldsymbol{x}}^{j,-j}_{i}[k+1] $ by (\ref{dis_sec_x}) and (\ref{dis_sec_xneg}) while  $\iota_{\{1\}}(i)=0 $ and $ w^{j,l}_{i,1}=0 $; $ \lambda^j_i[k+1] $ by (\ref{dis_sec_lambda}).}	
\end{algorithm}
 \par For cluster $ j $, we assume that the information of all agents (strategies $ x^j $ and gradients $ v^j=[v^j_1,\dots,v^j_{n_j}] $) are shared by a virtual center before and after iteration procedure. Furthermore, we assume that the information of agents outside this cluster is unknowable, and each agent can only replace the true value by prediction. Meanwhile, the estimation value $ \hat{\boldsymbol{x}}^{j,-j}_i $, two multiplier variables $ \boldsymbol{\lambda}^j $, $ \mu ^j $ and the auxiliary variable $ z^j $ need to be transmitted through the two undirected graphs and finally achieve the same result as the full decision information. When Algorithm \ref{algorithm} starts, agent $ i $ in cluster $ j $ needs to update
$ x_{i}^{j} $, $ \hat{\boldsymbol{x}}_{i}^{j,-j}  $, $ \lambda_{i}^{j} $, $z^{j} $,  $\mu^{j} $ from $ k $ to $ k' $, and it needs to update from $ i=1 $ to $ i=n_j $ (all agents in the cluster $ j $ need to be updated),  meanwhile, we use
$ w^{j,l}_{i,1} $ to make a distinction between leader agent and normal agent, since $ w^{j,l}_{i,1} $ will equal to $ 0 $, if it is not a leader agent.
The same as indicator function $ \iota_{\{1\}}(i) $, it equals to $ 0 $ while it is not a leader agent.
Then  agents  in cluster $ j+1 $ start to update from $k  $ to $ k' $ when all agents in cluster $ j $ are finished updating. Repeat the procedure from $ j=1 $ to $ j=m $ until all agents in the game finished updating. Then we start procedure $ k'\rightarrow k+1 $, and the whole process is the same with first one. Since the objective function of each agent is related to the strategies of all agents $ \boldsymbol{x} $, each agent needs the strategy information of all agents when updating its strategy.
\par \textit{Remark}:  Algorithm \ref{algorithm} is called distributed since every agent only knows its local information of cluster, the information of other agents outside the cluster can only be estimated.
However, if each agent's objective function is only related to strategies of  itself and agents outside the cluster
$ \theta^j_i(x^j_i,\boldsymbol{x}^{-j}) $ as in \cite{zeng2019}, \cite{deng2021generalized},
or only related to the strategy of itself and other clusters' leaders $\theta^j_i(x^j_i,\Gamma(\boldsymbol{x}^{-j}))  $  where $ \Gamma(\boldsymbol{x}^{-j})$  is the
stacked strategies of the leader agents of all the clusters except that of cluster $ j $ as in \cite{meng2020linear}, then there is no need for a virtual center in each cluster.
 And the $\tilde{v}^j_i $ in  Algorithm \ref{algorithm} becomes $ v^j_i $, Algorithm \ref{algorithm} changes from cluster distributed to agents distributed.
 The same virtual center setting can also be
 seen in \cite{zhou2021distributed}.
 The situation in this paper covers the above-mentioned situations.
\section{ALGORITHM DEVELOPMENT AND CONVERGENCE ANALYSIS} \label{section4}
\subsection{Algorithm Development} \label{subsubsectionA}
\par In this section, we show how the distributed algorithm for seeking the GNE under partial-decision information  is developed and gives the convergence analysis.
\par In compact notation, denote $ \boldsymbol{x}[k]=col((x^j_i[k])_{j\in \mathbb{N}_{m}^{*},i\in \mathbb{N}_{n_{j}}^{*}} )$,
$ \hat{\boldsymbol{x}}[k]=col((\hat{\boldsymbol{x}}^j_i[k])_{j\in \mathbb{N}_{m}^{*},i\in \mathbb{N}_{n_{j}}^{*}})$,
$ \boldsymbol{\mu} [k]=col((\mu^j[k])_{j\in  \mathbb{N}_{m}^{*}}) $,
$ \boldsymbol{\lambda}[k]=col((\lambda^j_i[k])_{j\in \mathbb{N}_{m}^{*},i\in \mathbb{N}_{n_{j}}^{*}}) $,
$ \boldsymbol{z}[k]=col((z^j[k])_{j\in  \mathbb{N}_{m}^{*}} )$,
$ \boldsymbol{b} =col((b^j)_{j\in \mathbb{N}_{m}^{*}})$,
and
we define $ \boldsymbol{L}\boldsymbol{\lambda}[k-k'] $, $ \boldsymbol{\Lambda}^T\boldsymbol{\mu}[k-k'] $, $ \boldsymbol{F}(\hat{\boldsymbol{x}})[k-k'] $, $ \hat{\boldsymbol{x}}[k-k'] $, $\boldsymbol{\lambda}[k-k']  $, $ \boldsymbol{\mu}[k-k'] $ the same as (\ref{procedure2}),
then $ k\rightarrow k' $ of  Algorithm \ref{algorithm} can be equivalently written in the compact form as follows:
	\small
		\begin{align}
			&\boldsymbol{x}\left[k^{\prime}\right]=\operatorname{Prox}_{H}^{\boldsymbol{\rho}_{R}}(\boldsymbol{x}[k]-\boldsymbol{\rho}_{R}(\boldsymbol{F}(\hat{\boldsymbol{x}}[k])+c R(\hat{\boldsymbol{L}}+\hat{\boldsymbol{L}}_{m}^{0}) \hat{\boldsymbol{x}}[k]\\
			&\qquad +\boldsymbol{L} \boldsymbol{\lambda}[k]+\boldsymbol{\Lambda}^{\mathrm{T}} \boldsymbol{\mu}[k])), \nonumber
		\end{align}
	\begin{align}
		&\hat{\boldsymbol{x}}^{-j}\left[k^{\prime}\right]=\hat{\boldsymbol{x}}^{-j}\left[k\right]-c \boldsymbol{\rho}_{s} S(\hat{\boldsymbol{L}}+\hat{\boldsymbol{L}}_{m}^{0}) \hat{\boldsymbol{x}}[k],
	\end{align}
	\begin{align}
		&\boldsymbol{\lambda}\left[k^{\prime}\right]=\boldsymbol{\lambda}[k]+\boldsymbol{\tau} \boldsymbol{L} R \hat{\boldsymbol{x}}[k],\label{centrallambda}
	\end{align}
	\begin{align}
	&\boldsymbol{z}\left[k^{\prime}\right]=\boldsymbol{z}[k]-\boldsymbol{\sigma} \boldsymbol{L}_{m}^{0} \boldsymbol{\mu}[k],\label{centralz}
\end{align}
	\begin{align}
	&\boldsymbol{\mu}\left[k^{\prime}\right]=P_{\mathbb{R}_{+}^{w m}}(\boldsymbol{\mu}[k]+\boldsymbol{v}(\boldsymbol{L}_{m}^{0} \boldsymbol{z}[k]+\boldsymbol{\Lambda} R \hat{\boldsymbol{x}}[k]-\boldsymbol{b})),\label{centralmu}
\end{align}
	\normalsize
and $ k'\rightarrow k+1 $ compact form is represent as:
\small
		\begin{align}
			&\boldsymbol{x}[k+1]=\boldsymbol{x}\left[k^{\prime}\right]+\boldsymbol{\rho}_{R} \boldsymbol{L} \boldsymbol{\lambda}\left[k-k^{\prime}\right]+\boldsymbol{\rho}_{R} \boldsymbol{\Lambda}^{T} 	\boldsymbol{\mu}\left[k-k^{\prime}\right]\\
			&\qquad +\boldsymbol{\rho}_{R} \boldsymbol{F}(\hat{\boldsymbol{x}})\left[k-k^{\prime}\right]+c\boldsymbol{\rho}_{R} R\left(\hat{\boldsymbol{L}}+\hat{\boldsymbol{L}}_{m}^{0}\right) \hat{\boldsymbol{x}}\left[k-k^{\prime}\right],
			\nonumber\\
			&\hat{\boldsymbol{x}}^{-j}[k+1]=\hat{\boldsymbol{x}}^{-\boldsymbol{j}}\left[k^{\prime}\right]+c \boldsymbol{\rho}_{s} 	S\left(\hat{\boldsymbol{L}}+\hat{\boldsymbol{L}}_{m}^{0}\right) \hat{\boldsymbol{x}}\left[k-k^{\prime}\right],\\
			&\boldsymbol{\lambda}[k+1]=\boldsymbol{\lambda}\left[k^{\prime}\right]-\boldsymbol{\tau}\boldsymbol{L} R \hat{\boldsymbol{x}}\left[k-k^{\prime}\right],\\
			&\boldsymbol{z}[k+1]=\boldsymbol{z}\left[k^{\prime}\right]+\boldsymbol{\sigma} \boldsymbol{L}_{m}^{0} \boldsymbol{\mu}\left[k-k^{\prime}\right],\\
			&\boldsymbol{\mu}[k+1]=\boldsymbol{\mu}\left[k^{\prime}\right]-\boldsymbol{\nu} \boldsymbol{L}_{m}^{0} 	\boldsymbol{z}\left[k-k^{\prime}\right]-\boldsymbol{\nu } \boldsymbol{\Lambda} R \hat{\boldsymbol{x}}\left[k-k^{\prime}\right],
		\end{align}
\normalsize where $ \boldsymbol{\rho}=diag(\rho^j_i \boldsymbol{I}_{q}) $,
$ \boldsymbol{\rho}_s=diag(\rho^j_i \boldsymbol{I}_{nq-\sum_{j=1}^{m}n_jn_jq_j})$,
$ \boldsymbol{\rho}_R=diag(\rho^j_i \boldsymbol{I}_{q_j}) $,
$ \boldsymbol{\tau} =diag(\tau^j_i)$,
$ \boldsymbol{\sigma}=diag(\sigma^j) $,
$ \boldsymbol{\nu}=diag(\nu^j) $,
$ \boldsymbol{L}=diag(\boldsymbol{L}^j)$,
$ \boldsymbol{L}^0_m=L^0_m \otimes \boldsymbol{I}_w $,
$ \hat{\boldsymbol{L}} = L\otimes \boldsymbol{I}_q $,
$ \hat{\boldsymbol{L}}^0_m= \hat{L}^0_m \otimes \boldsymbol{I}_q$,
when $ j\in \mathbb{N}_{m}^{*}, i\in \mathbb{N}_{n_{j}}^{*} $.
And $ \hat{L}^0_m $ is the expanded Laplacian matrix of $ \mathcal{G}_L $ which is given by
\small
\begin{equation*}
	\hat{L}^0_m=\begin{bmatrix}
		\textstyle \sum_{l=2}^m \boldsymbol{w}^{1,l}_{1,1}& \dots& -\boldsymbol{w}^{1,m}_{1,1}\\
		\vdots& \ddots&  \vdots&\\
		-\boldsymbol{w}^{m,1}_{1,1}& \ldots& 	\textstyle \sum_{l=1,l\neq m}^m \boldsymbol{w}^{m,l}_{1,1}
	\end{bmatrix},
\end{equation*}
\normalsize
and every item in $ \hat{L}^0_m $ is a matrix of particular size which is related to the corresponding item's position in $ L^0_m $,
e.g., for item $ -\boldsymbol{w}^{j,l}_{1,1} $, the corresponding item is  $ -w^{j,l}_{1,1} $, according to its position: row $ j $ column $ l $ ($ j\neq l $) in $ L^0_m $, it expands to size $ n_j \times n_l $ as
\small
\begin{equation*}
	-\boldsymbol{w}^{j,l}_{1,1}=\begin{bmatrix}
		-w^{j,l}_{1,1}& \dots& 0\\
		\vdots& \ddots&  \vdots&\\
		0& \ldots& 	0
	\end{bmatrix},
\end{equation*}
\normalsize
and for items on the diagonal, if the position is row $ j $  column $ j $, it expands to size $ n_j\times n_j $.
This  avoids the leader vector operating as \cite{zeng2019}.
\par Denote $ \varpi=col{(\hat{\boldsymbol{x}},\boldsymbol{z},\boldsymbol{\lambda},\boldsymbol{\mu})} \in \Omega$,
where $ \Omega:= \mathbb{R}^{nq}\times\mathbb{R}^{wm}\times\mathbb{R}^{q}\times\mathbb{R}^{wm}_{+} $.
Denote the block-diagonal matrix of the step sizes  and the skew symmetric respectively as
\begin{equation}\label{psi}
     \Psi =diag(\boldsymbol{\rho}^{-1}, \boldsymbol{\sigma}^{-1}, \boldsymbol{\tau}^{-1}, \boldsymbol{\nu}^{-1}),
\end{equation}
and
\begin{equation}\label{phi}
	\Phi=\begin{bmatrix}
		\boldsymbol{0}&  \boldsymbol{0}& R^{T} \boldsymbol{L}^{T} & R^{T} \boldsymbol{\Lambda}^{T} \\
		\boldsymbol{0} & \boldsymbol{0} & \boldsymbol{0} & \boldsymbol{L}_{m}^{0} \\
		-\boldsymbol{L} R & \boldsymbol{0} & \boldsymbol{0} & \boldsymbol{0} \\
		-\boldsymbol{\Lambda} R & -\boldsymbol{L}_{m}^{0} & \boldsymbol{0} & \boldsymbol{0}
	\end{bmatrix}.
\end{equation}
Define $ \mathcal{A}:\Omega \rightarrow \mathbb{R}^{nq+q+2wm}  $, $ \mathcal{B}: \Omega \rightarrow 2^{\mathbb{R}^{nq+q+2wm}} $ as
\small
\begin{equation}	\label{operators}
	\begin{aligned}
			&\mathcal{A}: \varpi \mapsto col(	R^{T} \boldsymbol{F}(\hat{\boldsymbol{x}})+c(\hat{\boldsymbol{L}}+\hat{\boldsymbol{L}}_{m}^{0}) \hat{\boldsymbol{x}},\mathbf{0}_{wm},\mathbf{0}_{q}, \boldsymbol{b})+ \Phi\varpi,\\
			&\mathcal{B}:\varpi  \mapsto R^TH(R\hat{\boldsymbol{x}})\times\boldsymbol{0}_{wm}\times\boldsymbol{0}_{q}\times N_{\mathbb{R}^{wm}_+}(\boldsymbol{\mu})	,
	\end{aligned}
\end{equation}
\normalsize
where $ R^TH(R\hat{\boldsymbol{x}}):=\{R^Tv |v \in H(R\hat{\boldsymbol{x}})\} $,
$ N_{\mathbb{R}^{wm}_+}(\boldsymbol{\mu})=\prod_{j=1}^{m}N_{\mathbb{R}^{w}_+}(\mu^j) $.
\begin{assumption}\label{assumption6psi}
$ \Psi^{-1}\succ0 $ and $ \|\Psi^{-1}\| <  1/{\ell_{\mathcal{A}}} $, where $\ell_{\mathcal{A}}  $ is the Lipschitz constant of $ \mathcal{A} $ (We prove the Lipschitz property in convergence analysis section.).
\end{assumption}
\par\textit{Remark}: Suppose Assumption \ref{assumption6psi} holds. Then the biggest step-size of  Algorithm \ref{algorithm} $ max((\rho^j_i,\tau^j_i,\sigma^j,\nu^j)_{j\in  \mathbb{N}_{m}^{*},i\in \mathbb{N}_{n_{j}}^{*}})\in (0,1/\ell_{\mathcal{A}}) $.
\begin{theorem}\label{lemma2}
	Suppose Assumption  \ref{assumption1}-\ref{assumption6psi}  hold. Let $ \varpi[k]=col{(\hat{\boldsymbol{x}}[k],\boldsymbol{z}[k],\boldsymbol{\lambda}[k],\boldsymbol{\mu}[k])} $, and $ \Psi$,   $ \Phi $, $\mathcal{A} $ and $ \mathcal{B} $,  as  in (\ref{psi}), (\ref{phi}), (\ref{operators}). Then  Algorithm \ref{algorithm} is equivalent to
	\begin{equation}
	      \boldsymbol{0} \in  \mathcal{A}(\varpi[k])+\mathcal{B}(\varpi[k]),
	\end{equation}  	
or
	\begin{equation}
	\varpi[k]:=\mathcal{T}\varpi[k],
  \end{equation}
where $ \mathcal{T}:= \Psi^{-1}\mathcal{A}+(\text{Id}-\Psi^{-1}\mathcal{A})\circ J_{\Psi^{-1}\mathcal{B}}\circ(\text{Id}-\Psi^{-1}\mathcal{A}) $, $J_{\Psi^{-1}\mathcal{B}}:=(\text{Id}+\Psi^{-1}\mathcal{B})^{-1} $. And any limit point of  Algorithm \ref{algorithm} is a zero of $ \mathcal{A}+\mathcal{B} $ and a fixed point of $ \mathcal{T} $.	
\end{theorem}
\par \textit{Proof}:	See appendix \ref{proofoflemma2} for details.	\hfill $\blacksquare$
\subsection{Convergence  Analysis} \label{subsubsectionB}
\par In this part, we show  the convergence of Algorithm \ref{algorithm}.

\begin{theorem}\label{theorem1}
	Suppose Assumption \ref{assumption1}-\ref{assumption6psi} hold. For any limit point
	$ \varpi^* =col(\hat{\boldsymbol{x}}^*,\boldsymbol{z}^*,\boldsymbol{\lambda}^*,\boldsymbol{\mu}^*) \in zer(\mathcal{A}+\mathcal{B})$ or
	$ \varpi^*=\mathcal{T}\varpi^* $, we have  $ \hat{\boldsymbol{x}}^* =\boldsymbol{1}_n \otimes \boldsymbol{x}^* $, $ \boldsymbol{\lambda}^* $, $ \boldsymbol{\mu}^* =\boldsymbol{1}_n \otimes 	\mu^* $ satisfy the KKT conditions (\ref{pKKT}), and $ \boldsymbol{x}^*$, $ \boldsymbol{\lambda}^* $, $ \mu^* $ satisfy the KKT condition  (\ref{vKKT} ). Moreover, $ \boldsymbol{x}^* $ solves the  (\ref{VI}), and is the v-GNE of the (\ref{min}) and (\ref{min2}).
\end{theorem}
\par \textit{Proof}:	See appendix \ref{proofoftheorem1} for details. 	\hfill $\blacksquare$

\par Combined with Lemma \ref{lemma2} and Theorem \ref{theorem1}, we can get that  Algorithm \ref{algorithm} comes from the transformation of a zero point of the sum of operators ($ \mathcal{A} $, $ \mathcal{B} $) or a fixed point of a mapping $ \mathcal{T} $. And when $ k\rightarrow \infty $, specific component of limit point $ \varpi^* $ of  Algorithm \ref{algorithm} satisfies the KKT conditions (\ref{VI}), (\ref{vKKT}), (\ref{pKKT}), and is a GNE of (\ref{min}) and (\ref{min2}) respectively.
\par In order to prove the maximal monotonicity of operator $ \mathcal{A} $ defined in (\ref{operators}), we show the $ c $-related restricted monotone property of  first line item of operator $ \mathcal{A} $  in the following \cite{pavel2019distributed}.
\begin{lemma} \label{lemmaofmonotone}
	Suppose Assumption \ref{assumption1}-\ref{assumption6psi} hold, let
	$ c_{min} =(({\kappa_0+\kappa})^2+4\eta\kappa)/4\eta(s_2(L)+s_2(L^0_m))$ and denote
\begin{equation}
	M=\begin{bmatrix}
	    \eta/n &   -({\kappa+\kappa_0})/2\sqrt{n}&\\
	    -({\kappa+\kappa_0})/2\sqrt{n}&    c(s_2(L)+s_2(L_m^0))-\kappa&
	\end{bmatrix},
\end{equation}
and if $ c\geq c_{min} $, then $ s_{min}(M)\geq0 $, $ R^{T} \boldsymbol{F}(\hat{\boldsymbol{x}})+c(\hat{\boldsymbol{L}}+\hat{\boldsymbol{L}}_{m}^{0}) \hat{\boldsymbol{x}} $  is restricted monotone.
\end{lemma}
\par \textit{Proof}:	See appendix \ref{proofoflemma3} for details. 	\hfill $\blacksquare$
\par In the following, by positive definite property of matrix of $ \Psi $ defined in  (\ref{psi}), we show the properties of $ \mathcal{A} $  and $ \mathcal{B} $.
\begin{lemma}
	Suppose Assumption \ref{assumption1}-\ref{assumption6psi} hold, and $ \Psi$,   $ \Phi $, $\mathcal{A} $, $ \mathcal{B} $,  as defined in (\ref{psi}), (\ref{phi}), (\ref{operators}) , $ c\geq c_{min} $ as in (\ref{lemmaofmonotone}).  Then operators $ \mathcal{A} $  and $ \mathcal{B} $ satisfy the following properties under Euclidean norm $ \|\cdot\|_2 $:
	\begin{enumerate}
		\item $\mathcal{A}$ and $ \mathcal{B} $ are maximally monotone.
		\item $\mathcal{A}$ is single-valued and Lipschitz continuous with parameter $ \ell_{\mathcal{A}} $.
	\end{enumerate}
\end{lemma}
\par \textit{Proof}:	See appendix \ref{proofoflemma4} for details. 	\hfill $\blacksquare$
\par Next, the properties of operators $\Psi^{-1} \mathcal{A} $ and  $ \Psi^{-1} \mathcal{B} $ are discussed under $ \Psi $-induced norm $ \|\cdot\|_{\Psi} $.
\begin{lemma}
	Suppose Assumption \ref{assumption1}-\ref{assumption6psi} hold, and $ \Psi$,   $ \Phi $, $\mathcal{A} $, $ \mathcal{B} $,  as defined in (\ref{psi}), (\ref{phi}), (\ref{operators}), $ c\geq c_{min} $ as in (\ref{lemmaofmonotone}),
	$  max((\rho^j_i,\tau^j_i,\sigma^j,\nu^j)_{j\in  \mathbb{N}_{m}^{*},i\in \mathbb{N}_{n_{j}}^{*}})\in (0,1/\ell_{\mathcal{A}}) $, then the operators $ \Psi^{-1}\mathcal{A} $ and $ \Psi^{-1}\mathcal{B} $ satisfy the following properties under the $ \Psi $-induced norm $\|\cdot\|_\Psi  $:
	\begin{enumerate}
		\item $\Psi^{-1} \mathcal{A}$ and $ \Psi^{-1} \mathcal{B} $ are maximally monotone.
		\item $\Psi^{-1} \mathcal{A}$ is single-valued and Lipschitz continuous with parameter $ \ell_{\mathcal{A}\Psi} $.
	\end{enumerate}
\end{lemma}
\par \textit{Proof}:	See appendix \ref{proofoflemma5} for details. 	\hfill $\blacksquare$
\par In the following, we show the convergence of  Algorithm \ref{algorithm}.
\begin{theorem}
	Suppose Assumption \ref{assumption1}-\ref{assumption6psi} hold, and $ \Psi$,   $ \Phi $, $\mathcal{A} $, $ \mathcal{B} $,  as defined in (\ref{psi}), (\ref{phi}), (\ref{operators}), $ c\geq c_{min} $ as in (\ref{lemmaofmonotone}),
	$ max((\rho^j_i,\tau^j_i,\sigma^j,\nu^j)_{j\in  \mathbb{N}_{m}^{*},i\in \mathbb{N}_{n_{j}}^{*}})\in (0,1/\ell_{\mathcal{A}}) $,
	the sequence
	$ \{\varpi[k]\}_{k\geq0} $ generated by  Algorithm \ref{algorithm}  is  Fej$\acute{e}$r monotone with respect to the fixed points set of $ \mathcal{T}$,
	 and converges to $ zer(\mathcal{A} +\mathcal{B}) $. Moreover, for each agent, like agent $ i $ in cluster $ j $,
	 the prime variable $ \{\hat{\boldsymbol{x}}^j_i[k]\}_{k\geq0} $ in $ \{\varpi^j_i[k]\}_{k\geq0} $ generated by  Algorithm \ref{algorithm} converges to  $ \hat{\boldsymbol{x}}^{j*}_i $, which is a component of  a v-GNE ($ \hat{\boldsymbol{x}}^{j*} $) of (\ref{min}) , and its local component,
	 $ \{\hat{x}^{j,j'}_{i,i'}[k]\}_{k\geq0} $ where $ j\in  \mathbb{N}_{m}^{*},i'\in  \mathbb{N}_{n_{j'}}^{*} $, converges to the corresponding component $ x^{j'*}_{i'} $ in $ \boldsymbol{x}^* $.
\end{theorem}
\par \textit{Proof}:	See appendix \ref{proofoftheorem2} for details. 	\hfill $\blacksquare$
 \section{Application in Energy Internet}\label{section5}
 \par In this section, a best energy generation strategy problem of EI is formulated as a GNP of multi-cluster game,
 and the proposed GNE seeking algorithm is implemented to solve this problem in a distributed manner.
 A numerical simulation example is carried out to validate the effectiveness of the algorithm.
\par  We assume that the EI is composed of several energy subnets while  each subnet is constituted with multiple prosumers who
own energy sources and can both produce and
consume energy as in \cite{etesami2018stochastic}.
 And every prosumer is an energy circle contains energy facilities (EFs), energy loads (ELs), energy storage devices (ESs).
 EFs are usually clean renewable facilities like wind turbines or solar
 photovoltaics and so forth, and also some conventional energy facilities like  fossil-fuel power stations.
 ELs are equipments that consumes energy which can be household appliances or machines used in factories.
 ERs are essential storage devices which are used to store energy since renewable energy facilities have intermittent and energy fluctuation.
 Suppose that all energy-subnets can generate enough power to support ELs in the circle, and all extra energy are stored in ESs which is used to compete for the utility markets like factories or charing station for plug-in cars.
 Then these energy subnets are treated as non-cooperative players competing for energy utility markets by adjusting the power generation strategy which needs the cooperation of all prosumers where the power price of every prosumer depends on the quantity of all power which the whole energy subnets can provide.

 In every energy subnet, prosumers are connected by both-way communication networks which  deliver estimations and some multipliers.
 Meanwhile, leader prosumer of subnets also communicate estimations and some multipliers in same structure network.
 To the purpose of unified management, we suppose prosumers must obey constraint rules of energy control center.
 Since every facility in the same energy subnet follows the same strategy, we assume all leaders satisfy constraint rules instead of all prosumers.

 Denote $ m $  the number of the energy subnets, and for energy subnet $ j $, there are $ n_j $ prosumers, and each prosumer has several facilities to generate power. All these energy subnets compete for $ w $ energy markets.
 We assume every prosumer has respective  ELs and ESs.
 $ x^j_i \in \mathbb{R}^{q_j}  $ denotes the quantity of stored energy of prosumer $ i $ in energy subnet $ j $.
 Similarly, $ \boldsymbol{x} $ is the power of all energy subnets can provide for competing the utility markets.
 We consider $ \boldsymbol{0}_{q_j}<x^j_i<r^j_i$, where $ r^j_i \in \mathbb{R}^{q_j}$ represents the maximum capacity of ESs of prosumer $ i $ in energy subnet $j  $.
After satisfying the power consumption by ELs in respective energy subnet, power stored by ESs is used to compete for the utilities.
The proper $ x^j_i $ is meaningful for the schedule planning of the maintenance or management of the whole energy-subnet.

 Suppose every prosumer $ i $ in energy subnet $ j $ has a energy allocation matrix $ T^j_i \in \mathbb{R}^{w\times q_j} $ to specify which utility market it participates in. Like \cite{pavel2019distributed} \cite{yi2019operator}, if the $ e $-th column of $ T^j_i  $ has its $ s $-th element as $ 1 $ if and only if prosumer generate $ (x^j_i)_e  $ power to utility $ U_s $, and all other elements are 0, where $ (x^j_i)_e  $ stands for $ e $-th element of $ x^j_i  $. Therefore, we denote $ T^j=[T^j_1 ,\dots, T^j_{n_j}] $, $ T=[T^1,\dots, T^m] $.

 Assume that every prosumer has a strong convex quadratic  production payoff function $ c^j_i (x^j_i)=(x^j_i)^TQ^j_ix^j_i+(q^j_i)^Tx^j_i-oi $, where $ Q^j_i \succ 0 $, $ o>0 $ and $ q^j_i \in \mathbb{R}^{q_j} $.
 Since all prosumers are involved for competing for energy utilities, then the demand function of utility $ s $ is given as
 $ (p^j_i)_s(\boldsymbol{x})=(p^j_i)_s-(d^j_i)_s(T\boldsymbol{x})_s $, where $ (p^j_i)_s, (d^j_i)_s>0 $. Then the objective function of prosumer $ i $ in energy subnet $ j $ is defined as   $f^j_i(x^j,\boldsymbol{x}^{-j})  =c^j_i(x^j_i)-((p^j_i)-d^j_i T\boldsymbol{x})^T T^j_ix^j_i$, and the objective function of energy subnet $ j $ is defined as
 $ f^j(x^j,\boldsymbol{x}^{-j})=\sum_{i=1}^{n_j} f^j_i(x^j,\boldsymbol{x}^{-j})$. And all leaders must satisfy a coupled constraint
 $ \sum_{j=1}^{m}A^{j} x_{1}^{j} \leq \sum_{j=1}^{m}b^{j} $, where $ A^j\in \mathbb{R}^{w\times q_j } $ stands for constraint matrix, and $ b^j\in \mathbb{R}_{+}^{w}  $ .

 Then we consider partial-decision scenario, i.e., every prosumer can obtain the real strategies information in energy subnet, however, it can't easily obtain the information outside the energy subnet.
 Firstly, we denote $ \hat{\boldsymbol{x}}^j_i $ the estimation of agent $ i $ in cluster $ j $,
 and  $ R^j_i $ the  matrix defined as (\ref{RS}), i.e., $ R^j_i \hat{\boldsymbol{x}}^j_i =x^j_i $.
Moreover, $ R^j=diag((R^j_i)_{i\in \mathbb{N}^*_{n_j}}) $, $ R=diag((R^j)_{j\in \mathbb{N}^*_{m}}) $,
$ R^j_s=\boldsymbol{1}^{n_j} \otimes R^j $,
$ R_s=diag((R^j_s)_{j\in \mathbb{N}^*_{m}}) $.
 Then the objective function is rewritten as $ f^j_i(x^j,\hat{\boldsymbol{x}}^{j,-j}_i)=(x^j_i)^TQ^j_ix^j_i+(q^j_i)^Tx^j_i-oi-(p^j_i)^T T^j_ix^j_i
 +\sum_{j'=1}^{m} \sum_{i'=1}^{n_{j'}} (\hat{x}^{j,j'}_{i,i'})^T (T^{j'}_{i'})^T (d^j_{i})^T T^j_i x^j_i  $,
 where $\hat{x}^{j,j'}_{i,i'}  $  stands for the estimation of prosumer $ i $ in energy subnet $ j $ to prosumer $ i' $ in energy subnet $ j' $.
 Since $ \boldsymbol{F}(\hat{\boldsymbol{x}})=col((\nabla_{x^{j}}f^j(x^j,\hat{\boldsymbol{x}}^{j,-j}))_{j\in \mathbb{N}_{m}^{*}} )$,
 for the ease of notation, we denote
 $ \bar{T}=diag(T^1_1,\dots,T^m_{n_m}) $,
$ \varXi:=\sum_{j=1}^{m}n_jn_jq_j $,
 $ \tilde{T}=diag(T^1,\dots,T^2,\dots,T^3,\dots)\in \mathbb{R}^{nw\times \varXi} $, and
 $ \tilde{\boldsymbol{T}}=diag({T,\dots,T})\in \mathbb{R}^{n w\times nq}  $,
 $ \boldsymbol{Q}=diag((Q^j_i)_{j\in \mathbb{N}_{m}^{*},i\in \mathbb{N}_{n_{j}}^{*}}) $,
 $ D=diag((d^j_i)_{j\in \mathbb{N}_{m}^{*},i\in \mathbb{N}_{n_{j}}^{*}}) \in \mathbb{R}^{nw\times nw} $,
$ p^j=col(p^j_i)_{i\in \mathbb{N}_{n_{j}}^{*}}) $, $ \boldsymbol{p}=col((p^j)_{j\in \mathbb{N}_{m}^{*}} )$, and $ \boldsymbol{q} $, $ \boldsymbol{r} $, $ \boldsymbol{b} $ are defined similarly.
 then $ \boldsymbol{F}(\hat{\boldsymbol{x}})=2\boldsymbol{Q}^T+\boldsymbol{q}-\bar{T}^T\boldsymbol{p}+\bar{T}^TD(\tilde{T}R_s\hat{\boldsymbol{x}}+\tilde{\boldsymbol{T}}\hat{\boldsymbol{x}}) $.
\par We consider the EI consisted of 3 energy subnets, as shown in Fig.2, where Energy subnets-1 has 4 prosumers, Energy subnets-2 has 2, Energy subnets-3 has 3, and they are competing for 2 utilities. All prosumers are connected by both-way communication networks.
\begin{figure}[htbp]
	\centering
	\includegraphics[width=3.5in]{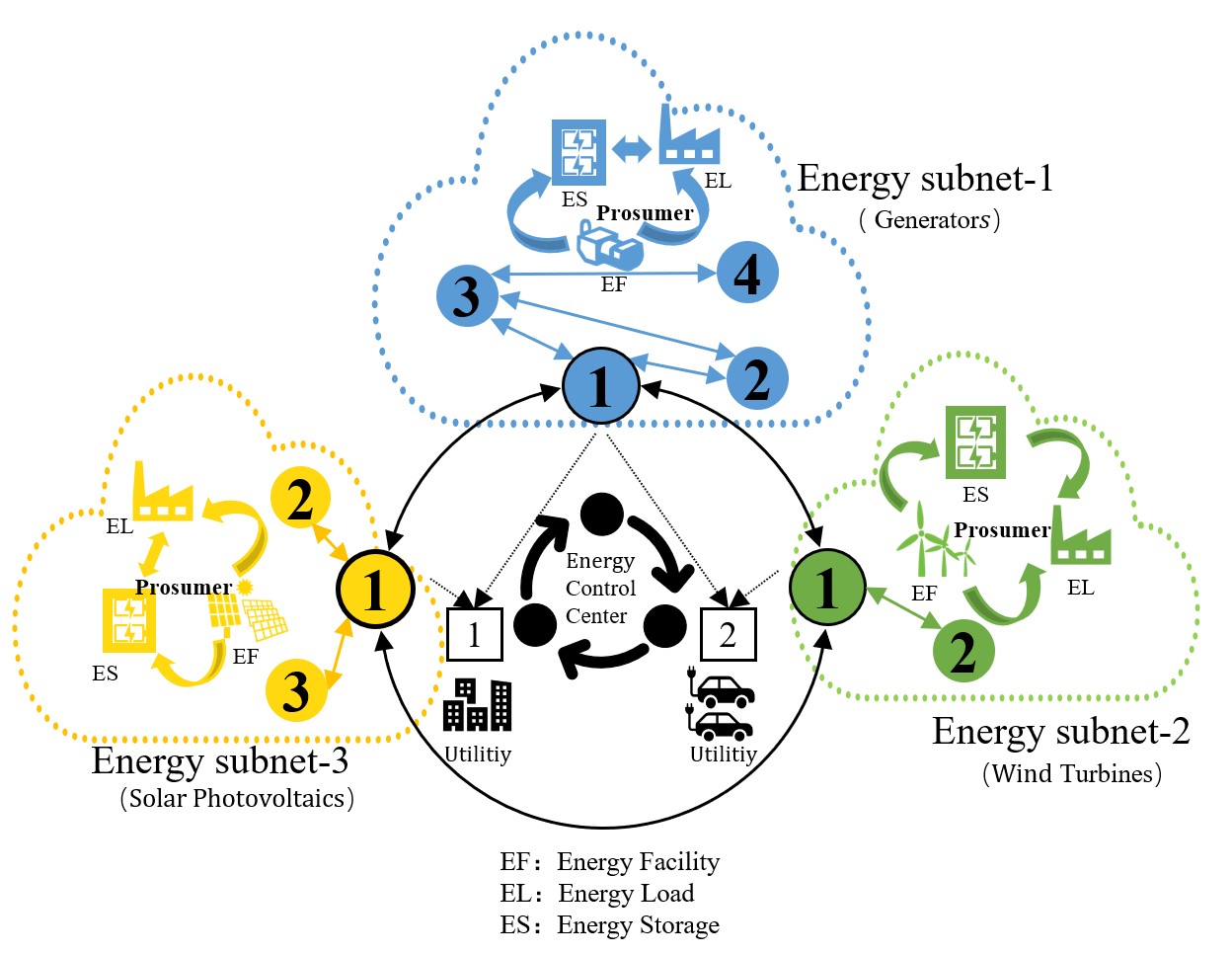}\\
	{Fig.2. Multi-cluster game of EI: Energy subnets-1(2,3) represent clusters, prosumers represent agents,
		and  each prosumer has EF (like wind turbines), EL and ES. }
\end{figure}
\par  And leader prosumers are required to satisfy constraints proposed by energy control center. In this setting, we assume $ r^j_i $ is randomly drawn from $ (5,10) $, and the same with $ b^j \in (1,2)$, $ p^j_i\in(10,20) $, $ d^j_i\in(1,3) $,
$ q^j_i\in(1,2) $, $ o=3 $, $ Q^j_i $ is diagonal with its entries from $ (1,8) $. And set $ c=70 $, $ \lambda=\tau=\sigma =\nu=0.002 $,
then the trajectories, including $ x^j_i $, $ \frac{1}{n_j-1}\sum_{i=2}^{n_j}\|x^j_1-x^j_i\|  $ , $ \frac{1}{n_j}\sum_{i=1}^{n_j}\|x^j_i-x^{j*}_i\| $ and $  \sum_{j=1,j\neq 3}^{3}\sum_{i=1}^{n_j}\|\hat{x}^{j,3}_{i,3}-x^{3*}_3\|  $ are shown in Fig.3, Fig.4 and Fig.5.
The relative error
$ \frac{\|\hat{\boldsymbol{x}}[k]-\hat{\boldsymbol{x}}^*[k]\|}{\|\hat{\boldsymbol{x}}^*[k]\|} $
generated by Algorithm 1 is shown in Fig.6.


\begin{figure}[htbp]
	\centering
	\includegraphics[width=3.6in]{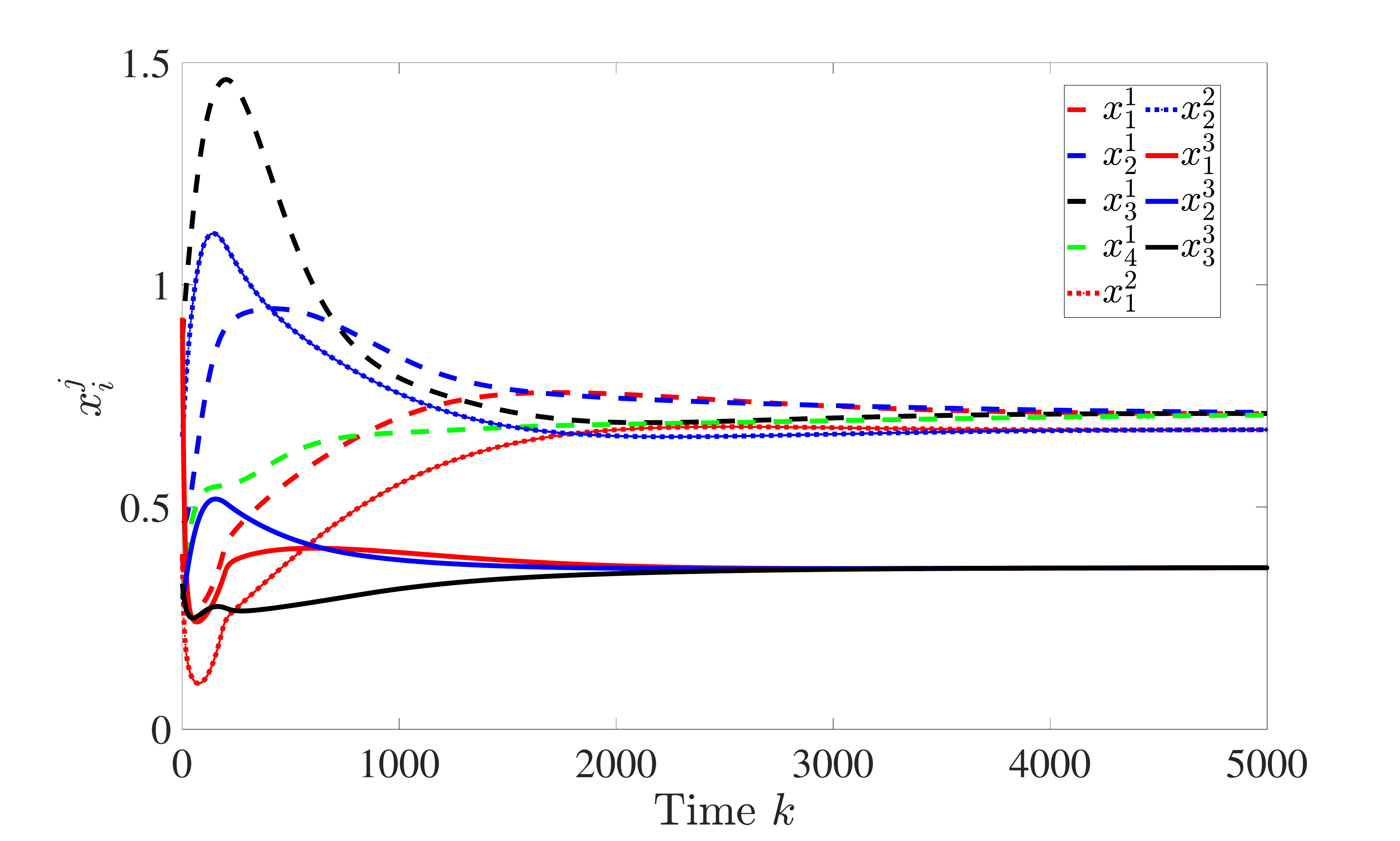}\\
	{Fig.3.  The trajectories of $x^j_i$, $ j\in \mathbb{N}_{3}^{*}$, $ i\in  \mathbb{N}_{n_{j}}^{*} $, where $ n_1=4 $, $ n_2=2 $, $ n_3=3 $. }
\end{figure}

\begin{figure}[htbp]
	\centering
	\includegraphics[width=3.6in]{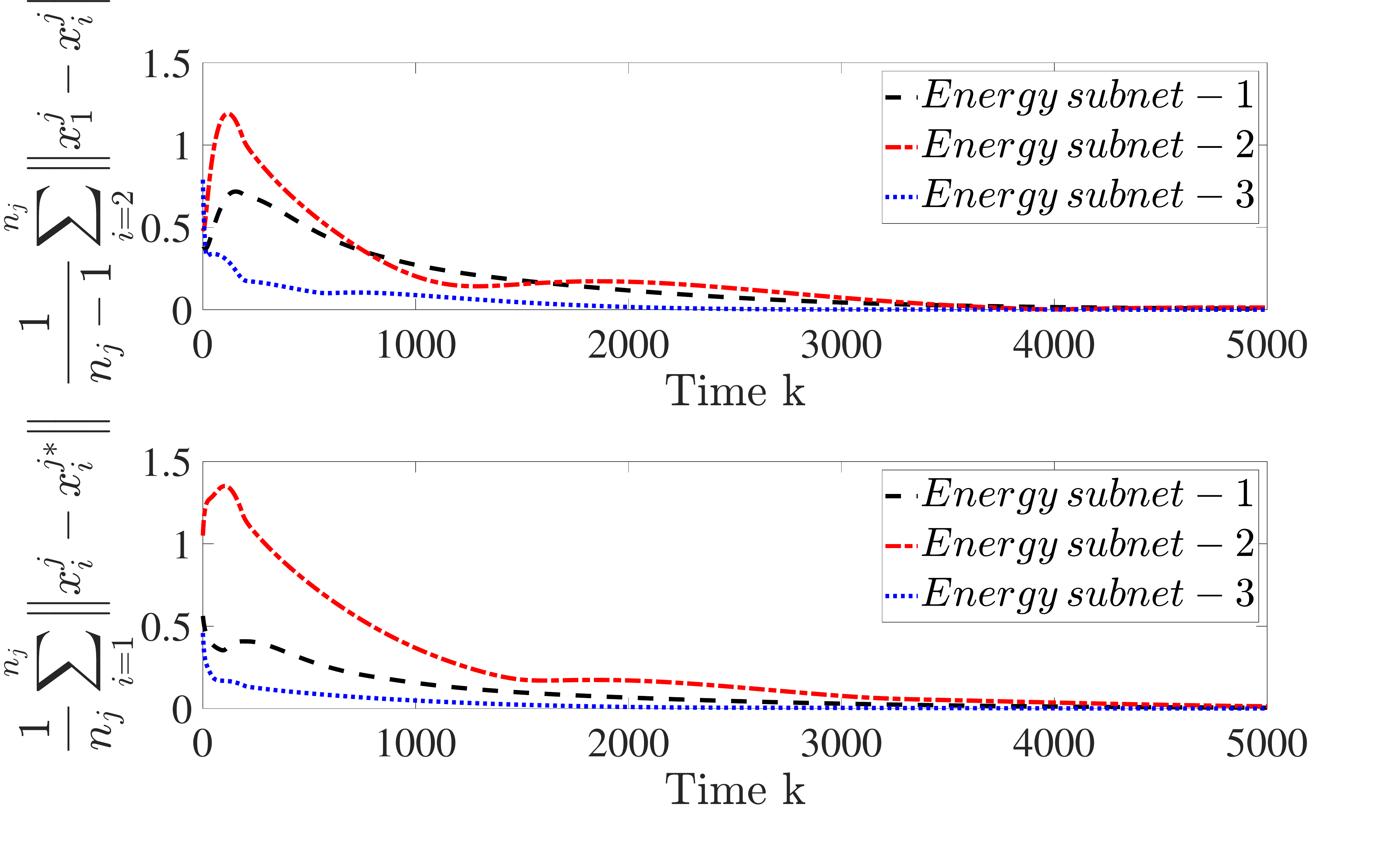}\\
	{Fig.4.  The trajectories of $ \frac{1}{n_j-1}\sum_{i=2}^{n_j}\|x^j_1-x^j_i\|  $ and $ \frac{1}{n_j}\sum_{i=1}^{n_j}\|x^j_i-x^{j*}_i\| $. }
\end{figure}

\begin{figure}[htbp]
	\centering
	\includegraphics[width=3.6in]{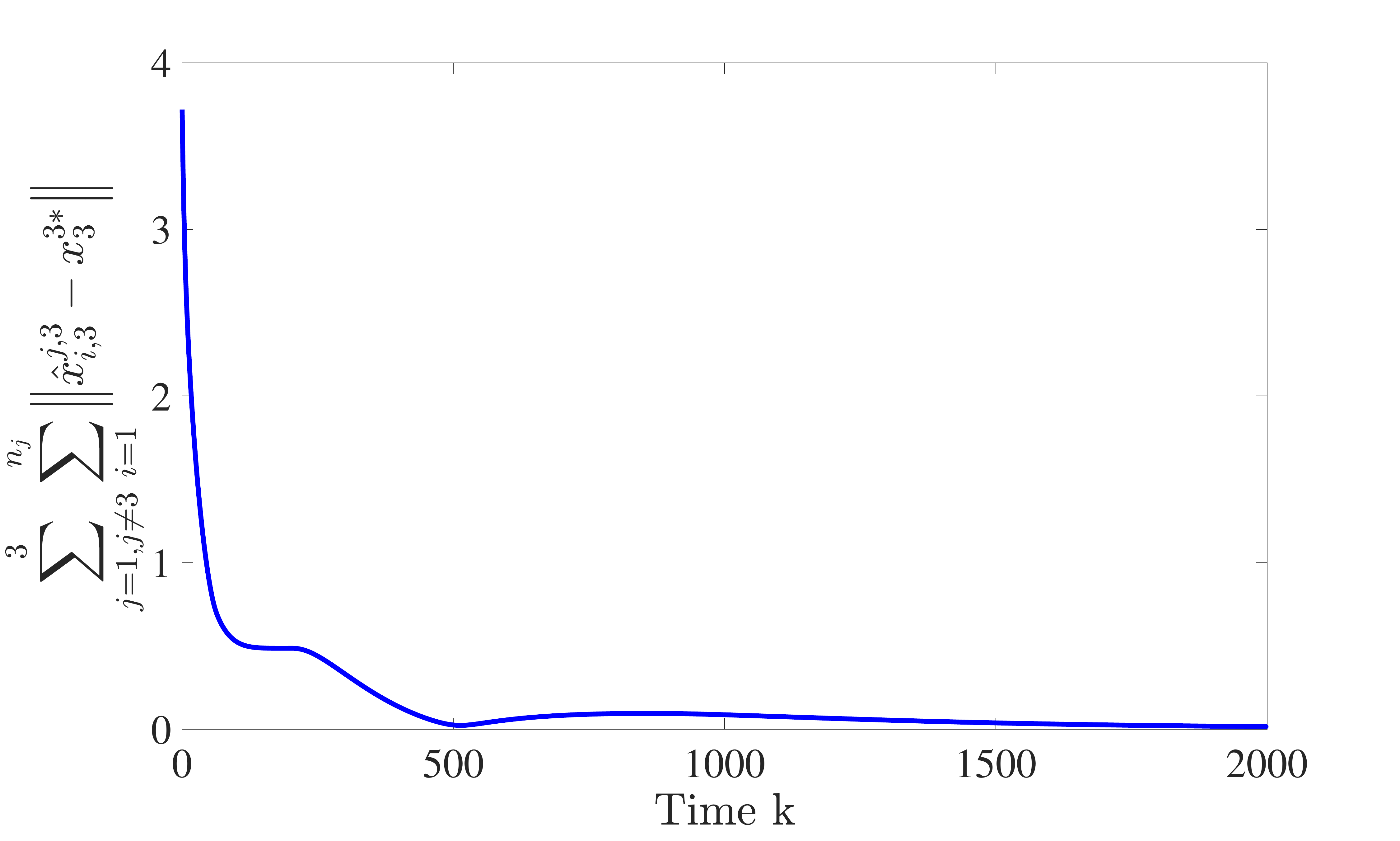}\\
	{Fig.5. The trajectories of $  \sum_{j=1,j\neq 3}^{3}\sum_{i=1}^{n_j}\|\hat{x}^{j,3}_{i,3}-x^{3*}_3\|  $}.
\end{figure}

\begin{figure}[!ht]
	\centering
	\includegraphics[width=3.6in]{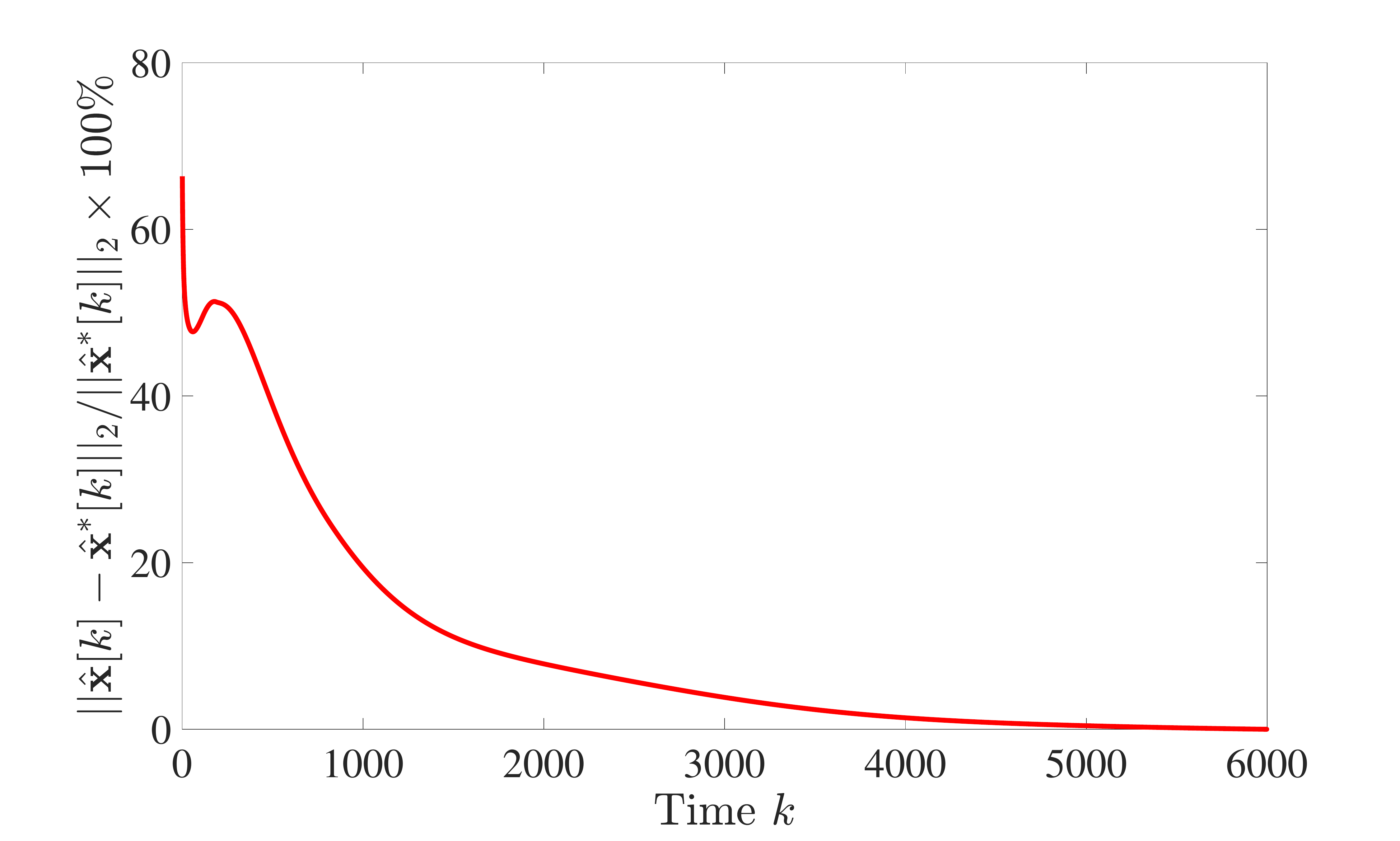}\\
	{Fig.6. Relative error generated by Algorithm 1: the trajectories of $ \frac{\|\hat{\boldsymbol{x}}[k]-\hat{\boldsymbol{x}}^*[k]\|}{\|\hat{\boldsymbol{x}}^*[k]\|} $. }
\end{figure}

\par  As shown in Fig.3 and Fig.4, the results satisfy the consistency constrain condition,
and show the convergence of Algorithm \ref{algorithm}.
Fig. 5 demonstrates that every prosumer's estimations to prosumer $ 3 $ in cluster $ 3 $ come to a consensus $ x^{3*}_3 $, which is a component of the Nash equilibrium.
Fig. 6 shows that the relative error  asymptotically decreases
to zero, implying that  $ \{\hat{\boldsymbol{x}}[k]\}_{(k\geq0)} $ asymptotically converges
to the Nash equilibrium.

\section{Conclusion}\label{section6}
\par  In this paper, we proposed a distributed GNE seeking algorithm designed by FBF iteration method of multi-cluster game with shared affine coupling constraints of leaders under partial-decision information scenario.
Convergence holds under restricted monotone of the extended pseudo-gradient mapping.
We discussed an example of multi-cluster game application in EI,
and the numerical simulation results shows the effectiveness of the algorithm.
%
%



%

\appendices
\section {Proofs} \label{appendix}
\textbf{Proof of Theorem 1}\label{proofoflemma2}: The analysis of  Algorithm \ref{algorithm} relies on the following equivalence:
\small
\begin{equation}\label{equi}
	\begin{aligned}
		&\boldsymbol{0} \in  \mathcal{A}(\varpi[k])+\mathcal{B}(\varpi[k]) \Leftrightarrow
		\varpi[k] =J_{\Psi^{-1}\mathcal{B}}(\text{Id}-\Psi^{-1}\mathcal{A})\varpi[k] \\
		&\Leftrightarrow (\text{Id}-\Psi^{-1}\mathcal{A})\varpi[k] =(\text{Id}-\Psi^{-1}\mathcal{A}) J_{\Psi^{-1}\mathcal{B}}(\text{Id}-\Psi^{-1}\mathcal{A})\varpi[k]\\
		&\Leftrightarrow \varpi[k]=((\text{Id}-\Psi^{-1}\mathcal{A})J_{\Psi^{-1}\mathcal{B}}(\text{Id}-\Psi^{-1}\mathcal{A})+\Psi^{-1}\mathcal{A})\varpi[k],
	\end{aligned}
\end{equation}
\normalsize
\par It is easy to see that fixed point  of
$ \mathcal{T}=((\text{Id}-\Psi^{-1}\mathcal{A})J_{\Psi^{-1}\mathcal{B}}(\text{Id}-\Psi^{-1}\mathcal{A})+\Psi^{-1}\mathcal{A}) $
and zero point of
$ \mathcal{A}+\mathcal{B} $ are equivalent. And we can generate two iteration equations from (\ref{equi}) as:
\small
\begin{align}
		\varpi[k'] &=J_{\Psi^{-1} \mathcal{B}}\left(\varpi[k]-\Psi^{-1} \mathcal{A} \varpi[k]\right) \label{iteration1},\\
		\varpi[k+1] &=\varpi[k']+\Psi^{-1}\left(\mathcal{A} \varpi[k]-\mathcal{A} \varpi[k']\right) \label{iteration2},
\end{align}
\normalsize
where (\ref{iteration1}) is denoted Forward-Backward procedure of  Algorithm \ref{algorithm}, and (\ref{iteration1}) is extra Forward procedure.
Note that (\ref{iteration1}) can be rewritten as: $ -\mathcal{A}\varpi[k]=\Psi(\varpi[k']-\varpi[k])+\mathcal{B}\varpi[k'] $,
it keeps us from calculating the inverse of $ \Psi $.
Then we use this equation and   (\ref{phi}), (\ref{operators}), (\ref{iteration1}), the first line  of the iteration equation is
\small
\begin{equation}	\label{firstline}
	\begin{aligned}
		&-(R^{T} \boldsymbol{F}(\hat{\boldsymbol{x}}[k])+c(\hat{\boldsymbol{L}}+\hat{\boldsymbol{L}}_{m}^{0})) \hat{\boldsymbol{x}}[k]+R^{T} \boldsymbol{L}^{T} \boldsymbol{\lambda}[k]+R^{T} \boldsymbol{\Lambda}^{T} \boldsymbol{\mu}[k])\\
		&\qquad =\boldsymbol{\rho}^{-1}(\hat{\boldsymbol{x}}[k^{\prime}]-\hat{\boldsymbol{x}}[k])+R^{T} H(R \hat{\boldsymbol{x}}[k^{\prime}]),
	\end{aligned}
\end{equation}
\normalsize 	
and we premultiply (\ref{firstline}) by $ R\boldsymbol{\rho} $ yields
\small
\begin{equation*}
	\begin{aligned}
	&-R \boldsymbol{\rho} R^{T} \boldsymbol{F}((\hat{\boldsymbol{x}}[k])) -c R \boldsymbol{\rho}\left(\hat{\boldsymbol{L}}+\hat{\boldsymbol{L}}_{m}^{0}\right)(\hat{\boldsymbol{x}}[k])-R \boldsymbol{\rho} R^{T} \boldsymbol{L}^{T} \boldsymbol{\lambda}[k]\\
	&-R \boldsymbol{\rho} R^{T} \boldsymbol{\Lambda}^{T} \boldsymbol{\mu}[k] = R\left(\left(\hat{\boldsymbol{x}}\left[k^{\prime}\right]-\hat{\boldsymbol{x}}[k]\right)\right)+R \boldsymbol{\rho} R^{T} H\left(R \hat{\boldsymbol{x}}\left[k^{\prime}\right]\right),
	\end{aligned}
\end{equation*}
\normalsize
since $ \boldsymbol{\rho}=diag(\rho^j_i \boldsymbol{I}_{q}) $,
$ \boldsymbol{\rho}_s=diag(\rho^j_i \boldsymbol{I}_{nq-\sum_{j=1}^{m}n_jn_jq_j})$,
$ \boldsymbol{\rho}_R=diag(\rho^j_i \boldsymbol{I}_{q_j}) $,
$ R^j=diag((R^j_i)_{i \in  \mathbb{N}_{n_{j}}^{*}})$, $ R=diag((R^j)_{j \in  \mathbb{N}_{m}^{*}})$,
$ S^j=diag((S^j_i)_{i \in  \mathbb{N}_{n_{j}}^{*}}) $, $ S^j=diag((S^j_i)_{j \in  \mathbb{N}_{m}^{*}})$, it follows that
$R \boldsymbol{\rho} R^{T}=\boldsymbol{\rho}_{R}$, $	R \boldsymbol{\rho}=\boldsymbol{\rho}_{R} R  $.
With $ (\text{Id}+\Psi \partial h)^{-1}=J_{\Psi\partial h}=Prox^{\Psi}_h $, $ R\hat{\boldsymbol{x}}[k]=\boldsymbol{x}[k] $, (\ref{firstline}) yields
	\small
\begin{align}
	&\boldsymbol{x}\left[k^{\prime}\right]=\operatorname{Prox}_{H}^{\boldsymbol{\rho}_{R}}(\boldsymbol{x}[k]-\boldsymbol{\rho}_{R}(\boldsymbol{F}(\hat{\boldsymbol{x}}[k])+c R(\hat{\boldsymbol{L}}+\hat{\boldsymbol{L}}_{m}^{0}) \hat{\boldsymbol{x}}[k] \nonumber\\
	&\qquad +\boldsymbol{L} \boldsymbol{\lambda}[k]+\boldsymbol{\Lambda}^{T} \boldsymbol{\mu}[k])). \nonumber
\end{align}
\par We  premultiply (\ref{firstline}) by $ S\boldsymbol{\rho} $ :
\small
\begin{equation*}
	\begin{aligned}
		-S \boldsymbol{\rho} R^{T} \boldsymbol{F}(\hat{\boldsymbol{x}}[k])-c S \boldsymbol{\rho} \hat{\boldsymbol{L}} \hat{\boldsymbol{x}}[k]-S \boldsymbol{\rho} R^{T} \boldsymbol{L} \lambda[k]-S \boldsymbol{\rho} R^{T} \boldsymbol{\Lambda}^{T} \boldsymbol{\mu}[k] \\
		=S\left(\hat{\boldsymbol{x}}\left[k^{\prime}\right]-\hat{\boldsymbol{x}}[k]\right)+S \boldsymbol{\rho} R^{T} H\left(R \hat{\boldsymbol{x}}\left[k^{\prime}\right]\right),
	\end{aligned}
\end{equation*}
 and with $ S\boldsymbol{\rho}=\boldsymbol{\rho}_s S $, $ S R^{T}=\mathbf{0}_{(n q-\sum_{j}^{m} n_{j} n_{j} q_{j}) \times(q)} $ yields
 \small
\[ \hat{\boldsymbol{x}}^{-j}[k+1]=\hat{\boldsymbol{x}}^{-\boldsymbol{j}}\left[k^{\prime}\right]+c \boldsymbol{\rho}_{s} 	S\left(\hat{\boldsymbol{L}}+\hat{\boldsymbol{L}}_{m}^{0}\right) \hat{\boldsymbol{x}}\left[k-k^{\prime}\right]. \]
\normalsize
\par The second line and the third line of the (\ref{iteration1}), it is easy to get (\ref{centrallambda}) and (\ref{centralz}). And with $ (\text{Id}+N_{\mathbb{R}^{wm}_+})^{-1}=P_{\mathbb{R}^{wm}_+} $, the fourth line yields (\ref{centralmu}).
\par Suppose Algorithm \ref{algorithm} has a limit point $ \varpi^* =col(\hat{\boldsymbol{x}}^*,\boldsymbol{z}^*,\boldsymbol{\lambda}^*,\boldsymbol{\mu}^*)$, by the continuity of the right-hand-side of
 Algorithm \ref{algorithm},  (\ref{equi}) holds. Then when $ k\rightarrow \infty $, $ \varpi[k]=\varpi^* $, i.e.,  any limit point of  Algorithm \ref{algorithm} is a zero of $ \mathcal{A}+\mathcal{B} $ and a fixed point of $ \mathcal{T} $ is satisfied.\\
\textbf{Proof of Theorem 2}:\label{proofoftheorem1}
Consider  $ \varpi^* =col(\hat{\boldsymbol{x}}^*,\boldsymbol{z}^*,\boldsymbol{\lambda}^*,\boldsymbol{\mu}^*) \in zer(\mathcal{A}+\mathcal{B})$,  with $ \Psi$,   $ \Phi $, $\mathcal{A} $ and $ \mathcal{B} $,  as defined in (\ref{psi}), (\ref{phi}), (\ref{operators}) , $ c\geq c_{min} $ as in (\ref{lemmaofmonotone}). Then it follows that
	\begin{align}
		&\mathbf{0}_{n q} \in R^{T} \boldsymbol{F}\left(\hat{\boldsymbol{x}}^{*}\right)+c(\hat{\boldsymbol{L}}+\hat{\boldsymbol{L}}_{m}^{0}) \hat{\boldsymbol{x}}^{*}+R^{T} \boldsymbol{L}^{T} \boldsymbol{\lambda}^{*}\nonumber \\
		&\qquad \qquad +R^{T} \boldsymbol{\Lambda}^{T}\boldsymbol{\mu}^{*}+R^{T} H\left(R \hat{\boldsymbol{x}}^{*}\right),\label{firstlineline}\\
		&\mathbf{0}_{w m}=\boldsymbol{L}_{m}^{0} \boldsymbol{\mu}^{*},\label{secondlineline} \\
		&\mathbf{0}_{n q}=-\boldsymbol{L} R \hat{\boldsymbol{x}}^{*},\label{thirdlineline} \\
		&\mathbf{0}_{w m} \in-\boldsymbol{L}_{m}^{0} \boldsymbol{z}^{*}-\boldsymbol{\Lambda} R \hat{\boldsymbol{x}}^{*}+\boldsymbol{b}+N_{\mathbb{R}_{+}^{w m}}\left(\boldsymbol{\mu}^{*}\right).\label{fourthlineline}
	\end{align}
\par In (\ref{firstlineline}), with $ \left(\mathbf{1}_{n}^{T} \otimes \boldsymbol{I}_{q}\right) \hat{\boldsymbol{L}}_{m}^{0}=\mathbf{0}_{q \times n q} $,
$ \left(\mathbf{1}_{n}^{T} \otimes \boldsymbol{I}_{q}\right) \hat{\boldsymbol{L}}=\left(\mathbf{1}_{n}^{T} \otimes \boldsymbol{I}_{q}\right)\left({L} \otimes \boldsymbol{I}_{q}\right)=\mathbf{1}_{n}^{T} L \otimes \boldsymbol{I}_{q}=\mathbf{0}_{q \times n q} $,
we premultiply $ \mathbf{1}_{n}^{T} \otimes \boldsymbol{I}_{q} $ to the first line yields
\begin{align}
	&\left(\mathbf{1}_{n}^{T} \otimes \boldsymbol{I}_{q}\right) \mathbf{0}_{n q} \in \left(\mathbf{1}_{n}^{T} \otimes \boldsymbol{I}_{q}\right) R^{T} \boldsymbol{F}\left(\hat{\boldsymbol{x}}^{*}\right)
	+\left(\mathbf{1}_{n}^{T} \otimes \boldsymbol{I}_{q}\right) R^{T} \boldsymbol{\Lambda}^{\mathrm{T}} \boldsymbol{\mu}^{*}\nonumber\\
	&+\left(\mathbf{1}_{n}^{T} \otimes \boldsymbol{I}_{q}\right) R^{T} \boldsymbol{L}^{T} \boldsymbol{\lambda}^{*}
	 +\left(\mathbf{1}_{n}^{T} \otimes \boldsymbol{I}_{q}\right) R^{T} H\left(R \hat{\boldsymbol{x}}^{*}\right), \nonumber
\end{align}
and by $ \left(\mathbf{1}_{n}^{T} \otimes \boldsymbol{I}_{q}\right) R^{T}=\boldsymbol{I}_{q} $, it follows that
\[
\mathbf{0}_{q} \in \boldsymbol{F}\left(\hat{\boldsymbol{x}}^{*}\right)+H\left(\boldsymbol{x}^{*}\right)+\boldsymbol{L}^{T} \boldsymbol{\lambda}^{*}+\boldsymbol{\Lambda}^{T} \boldsymbol{\mu}^{*}.
\]
\par If we premultiply $ R^T $, it follows that the first line of KKT condition ($ \ref{pKKT} $). And we can derive that $(\hat{\boldsymbol{L}}+\hat{\boldsymbol{L}}^0_m )\hat{\boldsymbol{x}}^*=\boldsymbol{0}_{nq} $.
By  $\hat{\boldsymbol{L} }=diag(\boldsymbol{L}^j)_{j\in  \mathbb{N}_{m}^{*}}$, $ \boldsymbol{L}^0_m=\hat{L}^0_m \otimes \boldsymbol{I}_q $, it follows that
$ \hat{\boldsymbol{x}}^* $ satisfies $\hat{\boldsymbol{x}}^{j*}_i =\hat{\boldsymbol{x}}^{j*}_{i'}  $, $\hat{\boldsymbol{x}}^{j*}_1 =\hat{\boldsymbol{x}}^{l*}_{1}  $, when $ i,i'\in  \mathbb{N}_{n_{j}}^{*} $ and  $ j,l\in  \mathbb{N}_{m}^{*} $.
Thus $ \hat{\boldsymbol{x}}^*\in \boldsymbol{E} $ the consensus subspace, then $\hat{\boldsymbol{x}}^*=\boldsymbol{1}_n \otimes \boldsymbol{x}^*$, then  $ \boldsymbol{F}\left(\hat{\boldsymbol{x}}^{*}\right)=\boldsymbol{F}\left(\mathbf{1}_{n} \otimes \boldsymbol{x}^{*}\right)=F\left(\boldsymbol{x}^{*}\right) $ holds, and by $ \boldsymbol{\Lambda}^T\boldsymbol{\mu }=\boldsymbol{A}^T\mu $, it follows that\[
\mathbf{0}_{q} \in F\left(\boldsymbol{x}^{*}\right)+H\left(\boldsymbol{x}^{*}\right)+\boldsymbol{L} \boldsymbol{\lambda}^{*}+\boldsymbol{A}^T \mu^{*},
\]
which is the first line of KKT condition (\ref{vKKT}).
\par In (\ref{secondlineline}), by $\boldsymbol{L}_{m}^{0}=L_{m}^{0} \otimes \boldsymbol{I}_{w} \in \mathbb{R}^{ w m \times w m}$, $\boldsymbol{\mu}^*=\mathbf{1}_{m} \otimes \mu^* \in \mathbb{R}^{w m} $, it follows that
\begin{align}
&\boldsymbol{L}_{m}^{0} \boldsymbol{\mu}^{*}=(L_{m}^{0} \otimes \boldsymbol{I}_{w}) \cdot (\mathbf{1}_{m} \otimes \mu) =L_{m}^{0} \mathbf{1}_{m} \otimes \boldsymbol{I}_{w} \mu\nonumber\\
&\qquad \quad =\mathbf{0}_{m} \otimes \boldsymbol{I}_{w} \mu=\mathbf{0}_{w m}.	\nonumber
\end{align}
\par In (\ref{thirdlineline}), $ R\hat{\boldsymbol{x}}^*=\boldsymbol{x}^* $, we can get $ \boldsymbol{L}\boldsymbol{x}^*=\boldsymbol{0}_q $. It is the third line of KKT condition ($ \ref{pKKT} $) and  ($ \ref{vKKT} $).
\par In (\ref{fourthlineline}), assuming that $ v_{1}, v_{2}, \ldots, v_{m} \in N_{\mathbb{R}_{+}^{w}}(\mu^*) $, using
$ \boldsymbol{L}_{m}^{0}=L_{m}^{0} \otimes \boldsymbol{I}_{w} $,
$ \boldsymbol{\Lambda}=diag((\boldsymbol{A}^j)_{j\in \mathbb{N}_{m}^{*}}) $,
$ \left(\mathbf{1}_{m}^{T} \otimes \boldsymbol{I}_{w}\right) \boldsymbol{L}_{m}^{0}=\mathbf{0}_{w \times w m} $,
$ \left(\mathbf{1}_{m}^{T} \otimes \boldsymbol{I}_{w}\right) \boldsymbol{\Lambda}=\boldsymbol{A} $,
$ \left(\mathbf{1}_{m}^{T} \otimes \boldsymbol{I}_{w}\right) \boldsymbol{b}=\sum_{j=1}^{m} b^{j}=b $,
we  premultiply $ \mathbf{1}_{m}^{T} \otimes \boldsymbol{I}_{w} $ to (37) it yields
\begin{equation*}
	\begin{aligned}
		&\left(\mathbf{1}_{m}^{T} \otimes \boldsymbol{I}_{w}\right) \mathbf{0}_{w m} \in-\left(\mathbf{1}_{m}^{T} \otimes \boldsymbol{I}_{w}\right) \boldsymbol{L}_{m}^{0} \boldsymbol{z}^*-\left(\mathbf{1}_{m}^{T} \otimes \boldsymbol{I}_{w}\right) \boldsymbol{\Lambda} R \hat{\boldsymbol{x}} \\
		&+\left(\mathbf{1}_{m}^{T}\otimes \boldsymbol{I}_{w}\right) \boldsymbol{b}+\left(\mathbf{1}_{m}^{T} \otimes \boldsymbol{I}_{w}\right) N_{\mathbb{R}_{+}^{w m}}(\boldsymbol{\mu}),
	\end{aligned}
\end{equation*}
and it can be equivalently written as $ \boldsymbol{0}_{w}=-\boldsymbol{A }\boldsymbol{x}^*+b+\sum_{j=1}^{m} v_{i} $,
which is
$ \mathbf{0}_{w} \in b-\boldsymbol{A} \boldsymbol{x}^{*}+N_{\mathbb{R}_{>0}^{w}}\left(\mu^{*}\right) $, the second line of KKT condition ($ \ref{pKKT} $) and ($ \ref{vKKT} $).\\
\textbf{Proof of Lemma 2}:\label{proofoflemma3} Define $ \boldsymbol{E}=\{\hat{\boldsymbol{x}}\in \mathbb{R}^{nq}|\hat{\boldsymbol{x}}^j_i =\hat{\boldsymbol{x}}^{j'}_{i'},  \forall j,j'\in  \mathbb{N}_{m}^{*}, i,i'\in  \mathbb{N}_{n_{j}}^{*}\}
=\{\hat{\boldsymbol{x}}=\boldsymbol{1}_n\otimes \boldsymbol{x}, \boldsymbol{x}\in \mathbb{R}^q \}$ as the estimation consensus subspace, $ \boldsymbol{E}^{\bot} $ its orthogonal complement with $ \mathbb{R}^{nq}=\boldsymbol{E}\oplus\boldsymbol{E}^{\bot} $.  Since $ \mathbb{R}^{nq}=\boldsymbol{E}\oplus\boldsymbol{E}^{\bot} $,
where
$ \boldsymbol{E}=\text{Null}(\hat{\boldsymbol{L}}+\hat{\boldsymbol{L}}^0_m)=\{\hat{\boldsymbol{x}}\in \mathbb{R}^{nq}|\hat{\boldsymbol{x}}^j_i =\hat{\boldsymbol{x}}^{j'}_{i'},  \forall j,j'\in  \mathbb{N}_{m}^{*}, i,i'\in  \mathbb{N}_{n_{j}}^{*}\}=\{\hat{\boldsymbol{x}}=\boldsymbol{1}_n\otimes \boldsymbol{x}, \boldsymbol{x}\in \mathbb{R}^q \} $ , $\boldsymbol{E}^{\bot}=\text{Null}(\hat{\boldsymbol{L}}+\hat{\boldsymbol{L}}^0_m)^{\bot} $ are the estimate consensus subspace and its orthogonal complement respectivly.
Any $ \hat{\boldsymbol{x}} $ can be decomposed as $\hat{\boldsymbol{x}}= \hat{\boldsymbol{x}}^\parallel+\hat{\boldsymbol{x}}^\bot  $,
where $\hat{\boldsymbol{x}}^\parallel \in  \boldsymbol{E}$ and
$\hat{\boldsymbol{x}}^\bot \in  \boldsymbol{E}^{\bot} $.
Thus $ \hat{\boldsymbol{x}}^\parallel = \boldsymbol{1}_n \otimes \boldsymbol{x} $ for  $ \boldsymbol{x}\in \mathbb{R}^q $.
And $ \text{min}_{\boldsymbol{x}^{\bot} \in \boldsymbol{E}^{\bot}}(\boldsymbol{x}^{\bot})^{T} \hat{\boldsymbol{L}} \boldsymbol{x}^{\bot}=s_{2}(L)\|\boldsymbol{x}^{\bot}\|^{2}$, since  $ s_{2}(\hat{\boldsymbol{L}})=s_{2}(L \otimes \boldsymbol{I}_{n})=s_{2}(L) $. We assume any $\hat{\boldsymbol{x}}'\in \boldsymbol{E}  $ can be rewritten as $ \hat{\boldsymbol{x}}'=\boldsymbol{1}_n\otimes\boldsymbol{x}'$, for any $ \boldsymbol{x}'\in \mathbb{R}^q $.
Then using $ \hat{\boldsymbol{x}}=\hat{\boldsymbol{x}}^{\parallel}+\hat{\boldsymbol{x}}^{\bot} $,
$ \boldsymbol{F}(\hat{\boldsymbol{x}}^{\parallel})=F(\boldsymbol{x}) $,
$ \boldsymbol{F}(\hat{\boldsymbol{x}}')=F(\boldsymbol{x}) $,
$ R \hat{\boldsymbol{x}}^{\|}=\boldsymbol{x} $,
$ R \hat{\boldsymbol{x}}'=\boldsymbol{x} $,
 $ (\hat{\boldsymbol{x}}-\hat{\boldsymbol{x}}^{\prime})^{T}(R^{T} \boldsymbol{F}(\hat{\boldsymbol{x}})-R^{T} \boldsymbol{F}(\hat{\boldsymbol{x}}^{\prime})+c(\hat{\boldsymbol{L}}+\hat{\boldsymbol{L}}_{m}^{0})(\hat{\boldsymbol{x}}-\hat{\boldsymbol{x}}^{\prime})) $ can be rewritten as the following \cite{pavel2019distributed},
\begin{align}
&(\boldsymbol{x}-\boldsymbol{x}^{\prime})^{T}(F(\hat{\boldsymbol{x}} )-F(\hat{\boldsymbol{x}}^{\|}))
+(\hat{\boldsymbol{x}}^{\perp})^{T}R^{T}(F(\hat{\boldsymbol{x}})-F(\hat{\boldsymbol{x}}^{\|})) \nonumber \\
&+(\boldsymbol{x}-\boldsymbol{x}^{\prime})^{T}(F(\boldsymbol{x})-F(\boldsymbol{x}^{\prime})) +(\hat{\boldsymbol{x}}^{\perp})^{T} R^{T}(F(\boldsymbol{x})-F(\boldsymbol{x}^{\prime})) \nonumber\\
&+c(\hat{\boldsymbol{x}}^{\perp})^{T} \hat{\boldsymbol{L}} \hat{\boldsymbol{x}}^{\perp}+c(\hat{\boldsymbol{x}}^{\perp})^{T} \hat{\boldsymbol{L}}_{m}^{0} \hat{\boldsymbol{x}}^{\perp},\nonumber
\end{align}
by Assumption \ref{assumption4strongmonotone}, \ref{assumption5extendgradient}, and
$ \|R^T\|=1 $,
$ \|\boldsymbol{x}^{\|}-\boldsymbol{x}^{\prime}\|=\|\mathbf{1}_{n} \otimes \boldsymbol{x}-\mathbf{1}_{n} \otimes \boldsymbol{x}^{\prime}\|=\|\mathbf{1}_{n}\otimes\left(\boldsymbol{x}-\boldsymbol{x}^{\prime}\right)\|=\|\mathbf{1}_{n}\|\|\boldsymbol{x}-\boldsymbol{x}^{\prime}\|
=\sqrt{n}\|\boldsymbol{x}-\boldsymbol{x}^{\prime}\| $, it follows that
\begin{equation*}
	\begin{aligned}
		&(\hat{\boldsymbol{x}}-\hat{\boldsymbol{x}}^{\prime})^{T}(R^{T} \boldsymbol{F}(\hat{\boldsymbol{x}})-R^{T} \boldsymbol{F}(\hat{\boldsymbol{x}}^{\prime})+c(\hat{\boldsymbol{L}}+\hat{\boldsymbol{L}}_{m}^{0})(\hat{\boldsymbol{x}}-\hat{\boldsymbol{x}}^{\prime}))\\
		&\geq -\kappa\|\boldsymbol{x}-\boldsymbol{x}^{\prime}\|\|\hat{\boldsymbol{x}}^{\perp}\|-\kappa\|\hat{\boldsymbol{x}}^{\perp}\|^{2}-\kappa_{0}\|\hat{\boldsymbol{x}}^{\perp}\|\|\boldsymbol{x}-\boldsymbol{x}^{\prime}\|\\
		&+c(s_{2}(L)+s_{2}(L_{m}^{0}))\|\hat{\boldsymbol{x}}^{\perp}\|^{2}+\eta\|\boldsymbol{x}-\boldsymbol{x}^{\prime}\|^{2}\\
		&\geq
		-\frac{\kappa}{\sqrt{n}}\|\hat{\boldsymbol{x}}^{\perp}\|\|\boldsymbol{x}^{\|}-\boldsymbol{x}^{\prime}\|-\frac{\kappa_{0}}{\sqrt{n}}\|\hat{\boldsymbol{x}}^{\perp}\|\|\boldsymbol{x}^{\|}-\boldsymbol{x}^{\prime}\| \\
		&\quad +c\left(s_{2}(L)+s_{2}\left(L_{m}^{0}\right)\right)\|\hat{\boldsymbol{x}}^{\perp}\|^{2}+\eta\|\boldsymbol{x}^{\|}-\boldsymbol{x}^{\prime}\|^{2} / n-\theta\|\hat{\boldsymbol{x}}^{\perp}\|^{2} \\
		& \geq
       \left[\begin{array}{c}
       	\|\boldsymbol{x}^{\|}-\boldsymbol{x}^{\prime}\| \\
       	\|\hat{\boldsymbol{x}}^{\perp}\|
       \end{array}\right]^{T} M\left[\begin{array}{c}
       	\|\boldsymbol{x}^{\|}-\boldsymbol{x}^{\prime}\| \\
       	\|\hat{\boldsymbol{x}}^{\perp}\|
       \end{array}\right],
   \end{aligned}
\end{equation*}
where
\begin{equation}
	M=\begin{bmatrix}
		\eta/n &   -({\kappa+\kappa_0})/2\sqrt{n}&\\
		-({\kappa+\kappa_0})/2\sqrt{n}&    c(s_2(L)+s_2(L_m^0))-\kappa&
	\end{bmatrix},
\end{equation}
thus, if $ c\geq (({\kappa_0+\kappa})^2+4\eta\kappa)/4\eta(s_2(L)+s_2(L^0_m))$, then $ M\succcurlyeq 0 $, which is
\begin{equation*}
(\hat{\boldsymbol{x}}-\hat{\boldsymbol{x}}^{\prime})^{T}(R^{T} \boldsymbol{F}(\hat{\boldsymbol{x}})-R^{T} \boldsymbol{F}(\hat{\boldsymbol{x}}^{\prime})+c(\hat{\boldsymbol{L}}+\hat{\boldsymbol{L}}_{m}^{0})(\hat{\boldsymbol{x}}-\hat{\boldsymbol{x}}^{\prime}))\geq 0.
\end{equation*}
\textbf{Proof of Lemma 3}:\label{proofoflemma4}
\par 1) Assume that $ \mathcal{A}= \mathcal{A}_1+\mathcal{A}_2$, where $\mathcal{A}_1= col(R^{T} \boldsymbol{F}(\hat{\boldsymbol{x}})+c(\hat{\boldsymbol{L}}+\hat{\boldsymbol{L}}_{m}^{0}) \hat{\boldsymbol{x}}, \mathbf{0}_{wm},\mathbf{0}_{q}, \boldsymbol{b})$,
	$ \mathcal{A}_2= \Phi\varpi $, $ \Phi $ in (\ref{phi}).
Since $ \Phi $ is a skew-symmetric matrix, i.e., $ \Phi^T=-\Phi $. As a single-valued operator, $ \Phi $ is maximally monotone (Ex.20.30,  \cite{bauschke2011convex}). And by Lemma \ref{lemmaofmonotone}, it follows that $ \mathcal{A}_1 $ is maximally monotone. Thus, $ \mathcal{A} $ is maximally monotone since the direct sum of maximally monotone
operators is maximally monotone ( Proposition 20.23, \cite{bauschke2011convex}).
$  N_{{\mathbb{R}^{wm}_+}} $ are maximally monotone as normal cones of closed convex sets, and  $ \boldsymbol{0}_{wm} $, $ \boldsymbol{0}_{nq} $  are also maximally monotone (Lemma 5, \cite{yi2019operator}). Meanwhile, function $ h^j_i $ is lower semi-continuous and convex as Assumption \ref{assumption1}. For each $ j\in  \mathbb{N}_{m}^{*} $, $ dom(h^j)=\Omega^j $ is a nonempty compact and convex set,
by $ H(R  \hat{\boldsymbol{x}} )=\partial_{x^{1}}h^1(x^1)\times \ldots \times \partial_{x^{m}}h^m(x^m) $, $\partial_{x^j}h^j(x^j)=\sum_{i=1}^{n_j}\partial_{x^{j}}h^j_i(x^j_i) $, it's easy to derived that the $ H(R\hat{\boldsymbol{x}}) $ which is the subdifferential of lower semi-continuous function, then it is maximally monotone. Thus, $ \mathcal{B} $ is maximally monotone.
\par 2) For operator $ \mathcal{A}_1 $, we can get
\begin{equation*}
	\begin{aligned}
		& \|\mathcal{A}_1\varpi-\mathcal{A}_1\varpi '\| \\
		&\leq
		\|R^{T} \boldsymbol{F}(\hat{\boldsymbol{x}})-R^{T} \boldsymbol{F}(\hat{\boldsymbol{x}}^{\prime})
		+c (\hat{\boldsymbol{L}}+\hat{\boldsymbol{L}}^0_m) (\hat{\boldsymbol{x}}-\hat{\boldsymbol{x}}^{\prime})\|	\\
		&\leq\|R^{T}\|\|\boldsymbol{F}(\hat{\boldsymbol{x}})-\boldsymbol{F}\left(\hat{\boldsymbol{x}}^{\prime}\right)\|+\|c (\hat{\boldsymbol{L}}+\hat{\boldsymbol{L}}^0_m)\|\|\hat{\boldsymbol{x}}-\hat{\boldsymbol{x}}^{\prime}\| \\
		& \leq \kappa\|R^{T}\|\|\hat{\boldsymbol{x}}-\hat{\boldsymbol{x}}^{\prime}\|+c(\|L\|+\|L^0_m\|)\|\hat{\boldsymbol{x}}-\hat{\boldsymbol{x}}^{\prime}\|\\
		&\leq (\kappa+cs_{n}(L)+cs_{m}(L^0_m)) \|\hat{\boldsymbol{x}}-\hat{\boldsymbol{x}}'\|,
	\end{aligned}
\end{equation*}
we denote $ \ell_{\mathcal{A}_1} =\kappa +cs_n(L)+s_m(L^0_m)$, then operator $ \mathcal{A} $ is $ \ell_{\mathcal{A}_1} $-Lipschitz.
\par For operator $ \mathcal{A}_2 $, it follows that
\begin{equation*}
	\begin{aligned}
		& \|\mathcal{A}_2\varpi -\mathcal{A}_2\varpi '\| \\
		&\leq\|R^{T} \boldsymbol{L}^{T} \boldsymbol{\lambda}-R^{T} \boldsymbol{L}^{T} \boldsymbol{\lambda}^{\prime}+R^{T} \boldsymbol{\Lambda}^{T} \boldsymbol{\mu}-R^{T} \boldsymbol{\Lambda}^{T} \boldsymbol{\mu}^{\prime}\|\\
		&\quad+\|\boldsymbol{L}_{m}^{0} \boldsymbol{\mu}-\boldsymbol{L}_{m}^{0} \boldsymbol{\mu}^{\prime}\|
		+\|-\boldsymbol{L} R \hat{\boldsymbol{x}}+\boldsymbol{L}R \hat{\boldsymbol{x}}^{\prime}\|\\
		&\quad+\|-\boldsymbol{L}_{m}^{0} \boldsymbol{z}+\boldsymbol{L}_{m}^{0} \boldsymbol{z}^{\prime}-\boldsymbol{\Lambda} R \hat{\boldsymbol{x}}+\boldsymbol{\Lambda} R \hat{\boldsymbol{x}}^{\prime}\|\\
		&\leq \|R^{T}\|\|\boldsymbol{L}^{T}\|\|\boldsymbol{\lambda}-\boldsymbol{\lambda}^{\prime}\|+\|R^{T}\|\|\boldsymbol{\Lambda}^{\mathrm{T}}\|\|\boldsymbol{\mu}-\boldsymbol{\mu}^{\prime}\| \\
		&\quad+\|\boldsymbol{L}_{m}^{0}\|\|\boldsymbol{\mu}-\boldsymbol{\mu}^{\prime}\|+\|\boldsymbol{L}\|\|R\|\|\hat{\boldsymbol{x}}-\hat{\boldsymbol{x}}^{\prime}\| \\
		&\quad+\|\boldsymbol{L}_{m}^{0}\|\|\boldsymbol{z}-\boldsymbol{z}^{\prime}\|+\|\boldsymbol{\Lambda}\|\|R\|\|\hat{\boldsymbol{x}}-\hat{\boldsymbol{x}}^{\prime}\|,
	\end{aligned}
\end{equation*}
since
$ \|\boldsymbol{\Lambda}^T\|=\|\boldsymbol{\Lambda}\|=max\{\delta_{max}(\boldsymbol{A}^1),\dots,\delta_{max}(\boldsymbol{A}^m)\}=max\{\delta_{max}(A^1),\dots,\delta_{max}(A^m)\}=\delta_{A^j}$.  $ \|\boldsymbol{L}^0_m=L^0_m \otimes \boldsymbol{I}^w \|=\|L^0_m\|=s_m(L^0_m)$, where $ s_m(L^0_m) $ is the maximum eigenvalue of $ L^0_m $, then it follows that
\begin{equation*}
\begin{aligned}
		 &\|\mathcal{A}_2\varpi -\mathcal{A}_2\varpi '\| \leq
		s_n(L)\|\boldsymbol{\lambda}-\boldsymbol{\lambda}^{\prime}\|
		+\delta_{A^j}\|\boldsymbol{\mu}-\boldsymbol{\mu}^{\prime}\| \\
		&\qquad\qquad+s_m(L^0_m) \|\boldsymbol{\mu}-\boldsymbol{\mu}^{\prime}\|
		+s_n(L)\|\hat{\boldsymbol{x}}-\hat{\boldsymbol{x}}^{\prime}\| \\
		&\qquad\qquad+s_m(L^0_m)\|\boldsymbol{z}-\boldsymbol{z}^{\prime}\|+\delta_{A^j}\|\hat{\boldsymbol{x}}-\hat{\boldsymbol{x}}^{\prime}\|,
\end{aligned}
\end{equation*}
let $ \ell_{\mathcal{A}_2}:=2s_n(L)+2s_m(L^0_m)+2\delta_{A^j} $, then $ \mathcal{A}_2 $ is $ \ell_{\mathcal{A}_2} $-Lipschitz. Thus,
$ \mathcal{A} $ is $(\ell_{\mathcal{A}}=\ell_{\mathcal{A}_1}+\ell_{\mathcal{A}_2} ) $-Lipschitz continuous (Prop. 20.23, \cite{bauschke2011convex}).
\par \textbf{Proof of Lemma 4}:\label{proofoflemma5}
\par 1) By the definition of monotone property, we need to prove
$ \langle  \Psi^{-1}\mathcal{A}(\varpi)-\Psi^{-1}\mathcal{A}(\varpi'),\varpi-\varpi'  \rangle_{\Psi} \geq 0$, by the monotone property of $ \mathcal{A} $ and definition of $ \Psi $-induced inner product, it follows that
\begin{equation*}
	\begin{aligned}
		\langle  \Psi^{-1}\mathcal{A}(\varpi)-\Psi^{-1}\mathcal{A}(\varpi')&,\varpi-\varpi'  \rangle_{\Psi} \\
		&= \langle  \mathcal{A}(\varpi)-\mathcal{A}(\varpi'),\varpi-\varpi' \rangle\geq0,\\
	\end{aligned}
\end{equation*}
and the same as $ \Psi^{-1}\mathcal{B} $.
\par 2) By (\ref{psi}) and the definition of $ \Psi $-induced norm, the $ \Psi  $-induced Lipschitz continuous property of $ \mathcal{A} $ is shown as follows :
\begin{equation*}
	\begin{aligned}
		&\|\Psi^{-1}\mathcal{A}(\varpi)-\Psi^{-1}\mathcal{A}(\varpi')\|_{\Psi}^2\\
		&=\langle\Psi\Psi^{-1}\mathcal{A}(\varpi)-\Psi\Psi^{-1}\mathcal{A}(\varpi'),\Psi^{-1}\mathcal{A}(\varpi)-\Psi^{-1}\mathcal{A}(\varpi')\rangle \\
		&\leq s_{max}(\Psi^{-1})\|\mathcal{A}(\varpi)-\mathcal{A}(\varpi')\|^2,
	\end{aligned}
\end{equation*}
since $ \Psi $ is the block-diagonal matrix and $ \Psi\succ 0 $, then the minimum and maximum eigenvalue of the $ \Psi $ and $ \Psi ^{-1} $ have satisfy $ s_{min}(\Psi)=1/s_{max}(\Psi^{-1}) $. And with Assumption \ref{assumption6psi} hold, i.e., $ \|\Psi^{-1}\|< 1/\ell_{\mathcal{A}} $, then $ \ell_{\mathcal{A}} < s_{min}(\Psi) $, by Lipschitz continuous property with $ \mathcal{A} $, it follows that \begin{equation*}
	\begin{aligned}
		&1/s_{min}(\Psi)\|\mathcal{A}(\varpi) -\mathcal{A}(\varpi')\|^2\leq {\ell_{\mathcal{A}}}^2/s_{min}(\Psi)\|\varpi-\varpi'\|^2\\
		& 	<s_{min}(\Psi)\|\varpi-\varpi'\|^2\leq  \|\varpi-\varpi'\|^2_{\Psi},
	\end{aligned}
\end{equation*}
thus, $ \|\Psi^{-1}\mathcal{A}(\varpi)-\Psi^{-1}\mathcal{A}(\varpi'))\|_{\Psi}^2 <  \|\varpi-\varpi'\|^2_{\Psi} $ holds, it means there exists a $ \ell_{\mathcal{A}\Psi} \in (0,1) $ such that $ \|\Psi^{-1}\mathcal{A}(\varpi)-\Psi^{-1}\mathcal{A}(\varpi')\|_{\Psi}^2 \leq \ell_{\mathcal{A}\Psi}  \|\varpi-\varpi'\|^2_{\Psi} $ holds, then 2) is proved.
\par \textbf{Proof of Theorem 3}:\label{proofoftheorem2} Suppose Assumption \ref{assumption5extendgradient} holds, there exists a fixed points set of $ \mathcal{T} $. Denote
$  \varpi^*\in fix(\mathcal{T}) $, $ \varpi\left[k^{\prime}\right]=J_{\Psi^{-1_{\mathcal{B}}}}\left(\varpi[k]-\Psi^{-1} \mathcal{A} \varpi[k]\right)$, $\varpi[k+1]=\mathcal{T}\varpi[k] $, for $ \varpi [k] \in H $ arbitrary. We can get
\begin{equation*}
	\begin{aligned}
		 &  \| \varpi[k+1]-\varpi^{*}  \|_{\Psi}^{2}\\
		 &\, = \|  \varpi[k+1]-\varpi [k^{\prime} ]+\varpi [k^{\prime} ]-\varpi[k]+\varpi[k]-\varpi^{*} \|_{\Psi}^{2}\\
		 &\,= \|\varpi[k+1]-\varpi [k^{\prime} ] \|_{\Psi}^{2}+ \|\varpi [k^{\prime} ]-\varpi[k] \|_{\Psi}^{2}+ \|\varpi[k]-\varpi^{*} \|_{\Psi}^{2} \\
		 &\qquad+2 \langle\varpi[k+1]-\varpi [k^{\prime} ], \varpi [k^{\prime} ]-\varpi^{*} \rangle_{\Psi}\\
		 &\qquad+2 \langle\varpi[k]-\varpi^{*}, \varpi [k^{\prime} ]-\varpi[k] \rangle_{\Psi},
	\end{aligned}
\end{equation*}
since $   \langle\varpi[k]-\varpi^{*}, \varpi  [k^{\prime}  ]-\varpi[k]  \rangle_{\Psi}=  \langle\varpi[k]-\varpi  [k^{\prime}  ]+\varpi  [k^{\prime}  ]-\varpi^{*}, \varpi  [k^{\prime}  ]-\varpi[k]  \rangle_{\Psi}=-\| \varpi[k]-\varpi  [k^{\prime}  ] \|^{2}+  \langle\varpi  [k^{\prime}  ]-\varpi^{*}, \varpi  [k^{\prime}  ]-\varpi[k]  \rangle_{\Psi} $, it follows that
\begin{equation*}
	\begin{aligned}
		 &\|\varpi[k+1]-\varpi^{*}  \|_{\Psi}^{2}=  \|\varpi[k]-\varpi^{*}  \|_{\Psi}^{2} +  \|\varpi  [k^{\prime}  ]-\varpi[k+1]  \|_{\Psi}^{2}
		 \\
		&-  \|\varpi[k]-\varpi  [k^{\prime}  ]  \|_{\Psi}^{2}+2  \langle\varpi  [k^{\prime}  ]-\varpi^{*}, \varpi[k+1]-\varpi[k]  \rangle_{\Psi}.
	\end{aligned}
\end{equation*}
\par Denote $ \varpi_{\mathcal{A}}[k]:=\mathcal{A} \varpi[k]$, $ \varpi_{\mathcal{A}}\left[k^{\prime}\right]:=\mathcal{A} \varpi\left[k^{\prime}\right]$,
$ \varpi_{\mathcal{B}}\left[k^{\prime}\right] \in \mathcal{B} \varpi\left[k^{\prime}\right] $, and
$ \varpi[k]=\varpi\left[k^{\prime}\right]+\Psi^{-1} \varpi_{\mathcal{B}}\left[k^{\prime}\right]+\Psi^{-1} \varpi_{\mathcal{A}}[k]$, $\varpi[k+1]=\varpi\left[k^{\prime}\right]+\Psi^{-1}\left(\varpi_{\mathcal{A}}[k]-\varpi_{\mathcal{A}}\left[k^{\prime}\right]\right) $, we can get
$ \Psi(\varpi[k]-\varpi[k+1])=\varpi_{\mathcal{B}}\left[k^{\prime}\right]+\varpi_{\mathcal{A}}\left[k^{\prime}\right] $. Moreover, denote
$ \varpi ^*_\mathcal{B} \in \mathcal{B}\varpi ^* $, $\varpi ^*_\mathcal{A} \in \mathcal{A}\varpi ^*  $, it follows that $ 0=\varpi^*_\mathcal{B}+\varpi^*_\mathcal{A} $ since $ 0\in \mathcal{B}\varpi^*+\mathcal{A}\varpi^* $.
Then $\langle\varpi  [k^{\prime}  ]-\varpi^{*}, \varpi[k+1]-\varpi[k]  \rangle_{\Psi}=  \langle\varpi  [k^{\prime}  ]-\varpi^{*}, \Psi(\varpi[k+1]-\varpi[k])  \rangle
	=  \langle-\varpi  [k^{\prime}  ]+\varpi^{*}, \varpi_{\mathcal{B}}  [k^{\prime}  ]-\varpi_{\mathcal{B}}^{*}-\varpi_{\mathcal{A}}^{*}+\varpi_{\mathcal{A}}  [k^{\prime}  ]  \rangle
	=  \langle-\varpi  [k^{\prime}  ]+\varpi^{*}, \varpi_{\mathcal{B}}  [k^{\prime}  ]-\varpi_{\mathcal{B}}^{*}  \rangle+  \langle-\varpi  [k^{\prime}  ]+\varpi^{*},-\varpi_{\mathcal{A}}^{*}+\varpi_{\mathcal{A}}  [k^{\prime}  ]  \rangle  $. Using the monotonicity of operator
	$ \mathcal{A} $ and $ \mathcal{B} $, it follows that $ \gamma =
	  -\langle-\varpi  [k^{\prime}  ]+\varpi^{*}, \varpi_{\mathcal{B}}  [k^{\prime}  ]-\varpi_{\mathcal{B}}^{*}  \rangle-  \langle-\varpi  [k^{\prime}  ]+\varpi^{*},-\varpi_{\mathcal{A}}^{*}+\varpi_{\mathcal{A}}  [k']  \rangle    \geq 0$.
	  \par  Furthermore,  $\|\varpi  [k^{\prime}  ]-\varpi[k+1]  \|_{\Psi}^{2}  =  \|\Psi^{-1}  (\varpi_{\mathcal{A}}  [k^{\prime}  ]-\varpi_{\mathcal{A}}[k]  )  \|_{\Psi}^{2}=  \langle\varpi_{\mathcal{A}}  [k^{\prime}  ]-\varpi_{\mathcal{A}}[k], \Psi^{-1}  (\varpi_{\mathcal{A}}  [k^{\prime}  ]-\varpi_{\mathcal{A}}[k]  )  \rangle
	  \leq s_{max}(\Psi^{-1})\|\varpi_{\mathcal{A}}  [k^{\prime}  ]-\varpi_{\mathcal{A}}[k]\|^2$, by the Lipschitz continuous property of $ \mathcal{A} $,
	  $ s_{max}(\Psi^{-1})\|\varpi_{\mathcal{A}}  [k^{\prime}  ]-\varpi_{\mathcal{A}}[k]\|^2
	  \leq \ell_{\mathcal{A}}^2s_{max}(\Psi^{-1})\|\varpi[k']-\varpi[k]\|^2 $.
	  Moreover, since $ \Psi\succ 0 $, by Rayleigh Quotient, $ s_{\min }(\Psi)  \|\varpi  [k^{\prime}  ]-\varpi[k]  \|^{2} \leq  \langle\Psi  (\varpi  [k^{\prime}  ]-\varpi[k]  ), \varpi  [k^{\prime}  ]-\varpi[k]  \rangle=  \|\varpi  [k^{\prime}  ]-\varpi[k]  \|_{\Psi}^{2} $.
	  Denote $\varepsilon:=1/{s_{max}(\Psi^{-1})}=s_{min}(\Psi)  $, thus, $   \|\varpi  [k^{\prime}  ]-\varpi[k+1]  \|_{\Psi}^{2} \leq  (\ell_{\mathcal{A}}\backslash \varepsilon)^{2}  \|\varpi  [k^{\prime}  ]-\varpi[k]  \|_{\Psi}^{2} $.
     \par  Therefore,
$      		 \|\varpi[k+1]-\varpi^{*}  \|_{\Psi}^{2}
     		 =  \|\varpi[k]-\varpi^{*}  \|_{\Psi}^{2}-  (1-  (\ell_{\mathcal{A}} / \varepsilon  )^{2}  )  \|\varpi  [k^{\prime}  ]-\varpi[k]  \|_{\Psi}^{2}-2 \gamma $
      holds. Since Assumption \ref{assumption6psi} holds, $ \{\varpi[k]\}_{k\geq 0}  $ is Fej$\acute{e}$r monotone and  $ \mathcal{T} $ is quasinonexpansive in $ \Psi $-induced inner product. Assume that $ \{\varpi [k']\}_{k\geq 0} $ is a converging subsequence with limit $ \varpi ^* $. It's easy to get from above analysis that $ \|\varpi[k^{\prime}]-\varpi[k]\|_{\Psi} \rightarrow 0, $ while $ k\rightarrow \infty $,
      and $  \|\varpi [k^{\prime} ]-\varpi[k] \|^{2}_{\Psi}= \langle\Psi (\varpi [k^{\prime} ]-\varpi[k] ), \varpi [k^{\prime} ]-\varpi[k] \rangle
      \quad= (\varpi [k^{\prime} ]-\varpi[k] )^{T} \Psi (\varpi [k^{\prime} ]-\varpi[k] ) $,  $ \Psi \succ 0 $ hold, it follows that
      $ \|\varpi[k^{\prime}]-\varpi[k]\| \rightarrow 0 \text { as } k \rightarrow \infty $, then by continuous of $ \mathcal{A} $,
      $ \|\varpi_{\mathcal{A}}[k^{\prime}]-\varpi_{\mathcal{A}}[k]\| \rightarrow 0 \text { as } k \rightarrow \infty $ holds.
       Since (31), we can get $ \varpi[k]-\Psi^{-1} \varpi_{ \mathcal{A}}[k] \in  \mathcal{B} \varpi\left[k^{\prime}\right]+\Psi^{-1}  \mathcal{B} \varpi\left[k^{\prime}\right] $, and we add $ \varpi _{\mathcal{A}}[k'] $ to each side, it yields $ \Psi (\varpi[k]-\varpi [k^{\prime} ] )+\varpi_{ \mathcal{A}} [k^{\prime} ]-\varpi_{ \mathcal{A}}[k] \in  \mathcal{A} \varpi [k^{\prime} ]+ \mathcal{B} \varpi [k^{\prime} ] $. Let $ k\rightarrow \infty $, then $ w_k=\Psi (\varpi[k]-\varpi [k^{\prime} ] )+\varpi_{ \mathcal{A}} [k^{\prime} ]-\varpi_{ \mathcal{A}}[k] \rightarrow 0 $, hence, $ 0\in \mathcal{B}\varpi^*+\mathcal{A}\varpi^* $ \cite{tseng2000modified}, \cite{franci2020half}, and it means all accumulation points are in
       $ fix(\mathcal{T}) $.
       \par We have proved when  $ k\rightarrow \infty $, $ \{\varpi [k]\}_{k\geq0} $ converges to $ \varpi^* $, so does its  component variable $ \hat{\boldsymbol{x}} $ which converges to $ \hat{\boldsymbol{x}}^* $.
       Since Theorem \ref{theorem1} has proved  $\hat{\boldsymbol{x}}\in \boldsymbol{E}  $, i.e.,  $ \hat{\boldsymbol{x}}=\boldsymbol{1}_n \otimes \boldsymbol{x}  $,
       for all $ j\in \mathbb{N}_{m}^{*}, i\in  \mathbb{N}_{n_{j}}^{*} $, estimate sequence
        $\{\hat{\boldsymbol{x}}^j_i[k]\}_{k\geq0} $ come consensus to $ \boldsymbol{x}^* $,
       and its local component
       $ \{\hat{x}^{j,j'}_{i,i'}[k]\}_{k\geq0} $, where $ j' \in  \mathbb{N}_{m}^{*},i'\in  \mathbb{N}_{n_{j'}}^{*} $ converges to the corresponding component
       $ x^{j'*}_{i'} $ in $ \boldsymbol{x}^* $.
	
%

%
%

\ifCLASSOPTIONcaptionsoff
  \newpage
\fi



%

\bibliographystyle{IEEEtran}
\bibliography{IEEEexample}
\end{document}